\documentclass[compsoc,journal]{IEEEtran}
\usepackage{cite}
\usepackage{subfiles}
\usepackage{amsmath,amsfonts}
\usepackage{graphicx}
\usepackage{booktabs}
\usepackage{tabularx}
\usepackage{multirow}
\usepackage{xcolor}
\usepackage{enumitem}
\usepackage{mathtools}
\usepackage{ragged2e}
\usepackage{stfloats}
\usepackage{color}
\usepackage{colortbl}
\usepackage[linesnumbered,ruled,vlined]{algorithm2e}
\usepackage{syntax}
\usepackage{caption}
\usepackage{nicefrac}
\usepackage{stmaryrd}
\usepackage{algpseudocode}
\usepackage{tabularx}
\usepackage{hyperref}
\usepackage[normalem]{ulem}
\usepackage[most]{tcolorbox}  
\usepackage[utf8]{inputenc}  
\usepackage[switch]{lineno}
\usepackage{lipsum}

\algnewcommand{\algorithmicand}{\textbf{ and }}
\algnewcommand{\algorithmicor}{\textbf{ or }}
\algnewcommand{\OR}{\algorithmicor}
\algnewcommand{\AND}{\algorithmicand}
\algnewcommand{\var}{\texttt}
\makeatletter 
\DeclareRobustCommand\newttfamily
{\not@math@alphabet\ttfamily\mathtt
  \fontfamily\ttdefault\selectfont\hyphenchar\font=`\-\relax}
\makeatother 
\DeclareTextFontCommand{\texttth}{\newttfamily}

\usepackage{listings}

\definecolor{javared}{rgb}{0.6,0,0} 
\definecolor{javagreen}{rgb}{0.25,0.5,0.35} 
\definecolor{javapurple}{rgb}{0.5,0,0.35} 
\definecolor{javadocblue}{rgb}{0.25,0.35,0.75} 
\definecolor{javagrey}{rgb}{0.46,0.45,0.48} 

\renewcommand{\algorithmiccomment}[1]{\bgroup\color{javagreen}{//#1}\egroup}
\newcommand\resq[1]{
\noindent 
\fcolorbox{green!40!black}{green!5}{\noindent 
 \parbox{0.98\columnwidth}{\noindent  #1}}\\
}
\lstdefinestyle{Spec}{
	language=Java, 
	keywordstyle=\color{javapurple}, 
	stringstyle=\color{javared},
	commentstyle=\color{javagreen},
	morecomment=[s][\color{javadocblue}]{/**}{*/},
	morecomment=[l][\color{javagrey}]{@},
	morecomment=[l][\color{javagrey}]{//@},
  morecomment=[l][\color{javagrey}]{/*@},
  morecomment=[l][\color{javagrey}]{*/},
	basicstyle=\ttfamily\footnotesize,
	breaklines=true,
	tabsize=2,
	frame=single,
	mathescape,
	numbers=left,
	xleftmargin=2.5em,
	xrightmargin=0.5em,
	frame=single,
	framexleftmargin=2em,
	morekeywords={,duration,simulationNodes,ms,Platform,CPU,memory,SimulationNode,platform,cloud,Cloud,IP,port,protocol,b,Simulator,username,password,quanta,step,Device,period,payload,speed,devices,locationIP, G, EdgeDevices,workload, inToOut, EdgeDevice,type, offsetRange,\%, pubTopic, subTopic,s, }
	escapeinside={*)} 
    columns=flexible,
	escapechar=?,
}

\lstdefinestyle{myCustomMatlabStyle}{
  language=Matlab,
  numbers=left,
  stepnumber=1,
  numbersep=10pt,
  tabsize=4,
  showspaces=false,
  showstringspaces=false
}

\usepackage{framed}
\definecolor{light-gray}{gray}{0.9}
\usepackage[framemethod=TikZ]{mdframed}
\mdfsetup{skipabove=5pt,skipbelow=3pt}
\usepackage{lipsum}
\mdfdefinestyle{MyFrame}{%
	linecolor=black,
	outerlinewidth=0.15pt,
	roundcorner=3pt,
	innertopmargin=2pt,
	innerbottommargin=2pt,
	innerrightmargin=4pt,
	innerleftmargin=4pt,
	backgroundcolor=light-gray}

\newcolumntype{K}[1]{>{\centering\arraybackslash}p{#1}}

\newcommand\lang{\textsf{IoTECS}}

\newcounter{commentnumber}

\begin{document}

\twocolumn

\title{A Lean Simulation Framework for Stress Testing \\ IoT Cloud Systems
}

\author{Jia Li,
Behrad Moeini,
Shiva Nejati, Mehrdad Sabetzadeh,
and Michael McCallen
\thanks{J. Li, B. Moeini, S. Nejati and M. Sabetzadeh are with University of Ottawa. \protect\\
E-mail: \{jli714, behrad.moeini, snejati, m.sabetzadeh\}@uottawa.ca}
\thanks{M. McCallen is with Cheetah Networks. \protect\\
E-mail: mccallen@cheetahnetworks.com}
}

\markboth{TRANSACTIONS ON SOFTWARE ENGINEERING}%
{Shell \MakeLowercase{\textit{et al.}}: Bare Demo of IEEEtran.cls for IEEE Journals}

\IEEEtitleabstractindextext{%
\begin{abstract}
The Internet of Things (IoT) connects a plethora of smart devices globally across various applications like smart cities, autonomous vehicles, and health monitoring. Simulation plays a key role in the testing of IoT systems, noting that field testing of a complete IoT product may be infeasible or prohibitively expensive. 
This paper addresses a specific yet important need in simulation-based testing for IoT: Stress testing of  cloud systems that are increasingly employed in IoT applications. Existing stress testing solutions for IoT demand significant computational resources, making them ill-suited and costly. We propose a lean simulation framework designed for IoT cloud stress testing. The framework enables efficient simulation of a large array of IoT and edge devices that communicate with the cloud. To facilitate simulation construction  for practitioners, we develop a \emph{domain-specific language (DSL)}, named \lang, for generating simulators from model-based specifications. We provide the syntax and semantics of \lang\ and implement \lang\ using Xtext and Xtend.  We assess simulators generated from \lang\ specifications  for stress testing two real-world systems: a cloud-based IoT monitoring system developed by our industry partner and an IoT-connected vehicle system.  Our empirical results indicate that simulators created using \lang: (1)~achieve best performance when configured with Docker containerization; (2)~effectively assess the service capacity of our case-study systems, and (3)~outperform industrial stress-testing baseline tools, JMeter and Locust, by a factor of  3.5 in terms of the number of IoT and edge devices they can simulate using identical hardware resources. 
To gain initial insights about the usefulness of \lang\ in practice, we interviewed two engineers from our industry partner who have firsthand experience with \lang. Feedback from these interviews suggests that \lang\ is effective in stress testing IoT cloud systems, saving significant time and effort.
\end{abstract}}

\IEEEdisplaynontitleabstractindextext
\maketitle


\section{Introduction}
\label{sec:intro}
\IEEEPARstart{I}{oT} envisions complex systems that interconnect large numbers of smart devices, embedded with sensors and actuators, through the Internet or other network technology~\cite{IoT17}. As IoT continues to find applications in various domains such as emergency response, smart cities, smart agriculture, and autonomous vehicles~\cite{IoT17}, providers of IoT cloud solutions are faced with the important question of how their systems will scale to handle the demand posed by vast arrays of IoT sensors, actuators, and gateways communicating with the cloud. Systematically answering this question requires \emph{stress testing}, which has to do with the analysis of how a given system behaves under extreme workloads and at the upper limits of its capacity~\cite{Chan:04}. Given the cost-prohibitive nature of building physical IoT testbeds, \emph{simulation} often emerges as the most practical approach for stress testing~\cite{Shin:20, Li:22, Li:23}.

\begin{figure}[t]
\begin{center}
    \includegraphics[width=.8\linewidth]{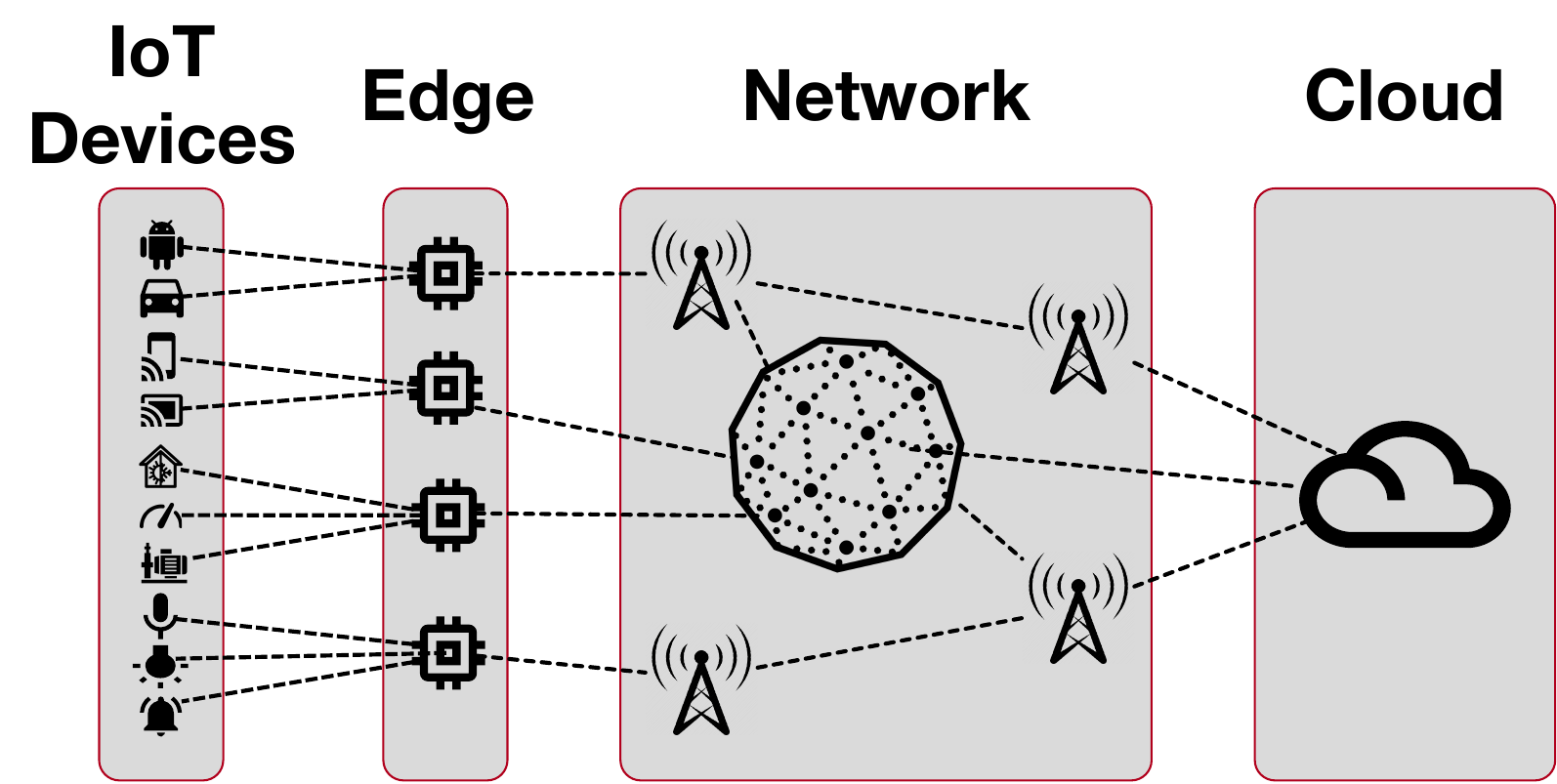}
    \end{center}
    \caption{A schematic overview of the common tiers in IoT systems.}
    \label{fig:edge-cloud}
\end{figure}

Figure~\ref{fig:edge-cloud} shows a schematic view representing the common tiers in IoT systems, encompassing the device, edge, network, and cloud layers. IoT simulators may target one or several of these tiers, such as simulating IoT sensors and actuators~\cite{Kertesz:19, AmazonIoT, NuvIoT}, IoT edge devices~\cite{Jha:20, Sonmez:17, Ramprasad:19}, and IoT networks~\cite{Kliazovich:12, Osterlind:06, boulis:11, Mininet}.
Among these simulators, those emulating the traffic load communicated between edge and cloud systems, known as \emph{edge-to-cloud simulators} \cite{Elazhary:19}, are applicable for stress testing IoT cloud systems.

There are a few IoT edge-to-cloud simulators in the literature~\cite{Ramprasad:19, Sonmez:17}; however, these simulators capture extensive details about IoT and edge devices, such as device locations, network protocols, battery capacities, and battery drainage rates. Due to their feature-rich nature, these simulators demand significant computational resources (CPU and memory) to simulate thousands of devices, making them unsuitable for  stress testing: Many IoT solutions providers, including our industry partner (see Section~\ref{subsec:partner}), rely on hosted services for simulation. The greater the resource demands of the simulator, the more challenging and costly it becomes to set up and run for stress testing, limiting the number of IoT and edge devices that can be simulated.

\subsection{Edge-to-Cloud Simulator Design} 
\label{subsec:features}

In this article, we develop a lean, purpose-built, edge-to-cloud simulation framework for stress testing IoT cloud systems. 
Our framework introduces the following novel design features to address scalability challenges in stress-testing of IoT cloud systems:

\textbf{F1.} \emph{IoT devices are represented symbolically.} Existing IoT simulators capture IoT devices as standalone processes or objects to model their detailed features~\cite{Ramprasad:19}, \cite{Kertesz:19}, \cite{Pflanzner:16}. In contrast, our simulation framework represents IoT devices \emph{symbolically}, focusing exclusively on capturing key IoT device properties relevant to stress testing: device payload and communication methods with edge devices.
This design feature, denoted by \textbf{F1}, results in more parsimonious and efficient simulators, helping address the \emph{run-time bloat} observed in existing edge-to-cloud simulators.

\textbf{F2.} \emph{Edge devices are augmented with configurable variables to account for the variability in start times of edge-device execution and the intervals between data transmissions from IoT devices.} These variables minimize bursts in edge-to-cloud data transmission, improving the operational capacity of simulators and enhancing the realism of simulations. This design feature, denoted by \textbf{F2}, leads to more consistent network bandwidth usage over time, helping mitigate \emph{bursty communication}.

\textbf{F3.} \emph{Edge devices are grouped into clusters using simulation nodes. These  nodes enable more scalable deployment of simulators.} Our simulation framework supports the grouping of simulated edge devices into different clusters, referred to as \emph{simulation nodes}, which can be executed directly on the native host machine, in independent containers and/or virtual machines, or in a combination of these environments. This design feature, denoted by \textbf{F3}, helps address the problem of \emph{scaling under limited computational resources}. Specifically, \textbf{F3} enables more efficient resource allocation, ensuring that each simulation node receives the necessary resources based on its workload. This approach further ensures isolation and stability; issues within one node are contained within its container, preventing system-wide impacts. While some current IoT simulators employ containerization and virtualization, our experiments in Sections~\ref{sec:usecase} and~\ref{sec:eval} indicate that none match our method in supporting a large number of IoT and edge devices under restricted CPU and memory resources.

\begin{figure}[t]
\begin{center}
    \includegraphics[width=.65\linewidth]{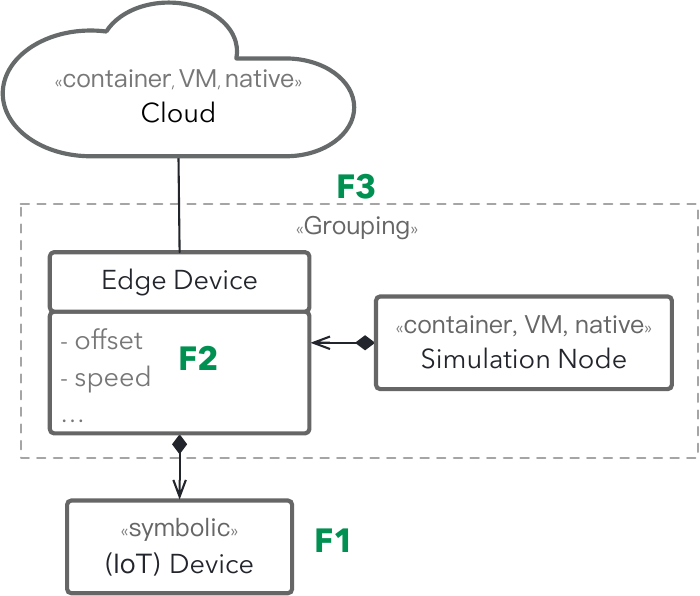}
    \end{center}
    \caption{Novel design features of our edge-to-cloud simulation framework.}~\label{fig:hierachical}
\end{figure}

Figure~\ref{fig:hierachical} provides an overview of the architecture of simulators in our framework, illustrating the three features described above. In our design, edge devices are grouped into simulation nodes. These nodes can run on a native operating system or use virtual machines or Docker containers with designated resources (\textbf{F3}). Each edge device connects to an array of IoT devices, which are represented by variables capturing the relevant properties of the IoT devices (\textbf{F1}). Unlike IoT devices, each edge device functions as an independent thread, communicating with the cloud. The method, data rate, and frequency of edge-to-cloud communication depend on the number and properties of the IoT devices, as well as the \emph{offset} and \emph{speed} variables (\textbf{F2}). Offset refers to the timing of edge-device executions, while speed denotes the rate at which edge devices send IoT-device data to the cloud.

To automate the creation of IoT edge-to-cloud simulators, our framework is supported by a domain-specific language (DSL), \lang\ (IoT Edge-to-Cloud Simulation Language). Specifications (models) written in \lang\ are automatically translatable into Java-based simulators. 
Note that while \lang\ specifications are not executable per se, the simulator code is
automatically and entirely derived from these specifications. The generated simulators are then executed to produce simulation results.

\subsection{Collaboration with Industry}\label{subsec:partner}

Our simulation framework is the outcome of a collaborative research initiative with Cheetah Networks (\url{https://cheetahnetworks.com}). Cheetah Networks is a Canadian company specializing in providing solutions for monitoring and optimizing the performance of IoT networks. Their platform employs  machine learning  to analyze data from IoT and edge devices, assisting network operators, businesses, and organizations in improving the reliability, efficiency, and security of their networks. The company's services cater to various industries, including mining, oil and gas, telecommunications, electric utilities, and smart cities, with clients spanning North America,  Europe, and the Middle East.

To ensure that rollouts of new services are not affected by scalability issues when released to clients, Cheetah Networks performs extensive stress testing prior to releases. Since building a testbed with thousands of actual IoT and edge devices is prohibitively expensive, the company uses simulation to mimic real-world loads on their cloud-based platform. Nevertheless, this approach has its own challenges: As we detail in Section~\ref{sec:usecase}, existing IoT simulators are not suitable for stress testing. Cheetah Networks has therefore relied on in-house testing tools to meet their stress testing needs. These in-house tools are expensive to maintain, require significant manual effort from the company's testing team to configure for different cloud services, are difficult to scale, and are prone to configuration errors. Moreover, these tools expose the testing team to intricate network programming, thereby significantly increasing the learning curve for new staff. The primary goal of our collaboration with Cheetah Networks has been to develop a simulation-based stress testing framework that helps address these challenges. Since its development, our simulation framework has been adopted by Cheetah Networks and, as of this writing, has been continuously applied for more than six months (without any involvement from the researchers) to stress-test some of the company's cloud-based services.

\subsection{Contributions}  
This article extends a previous conference paper~\cite{Li:22-2}, published at the 25th International Conference on Model Driven Engineering
Languages and Systems (MODELS~2022). Compared to our earlier publication, this current article offers major expansions in the following areas:

\begin{enumerate}
\item~We extend \lang\ to include support for the MQTT protocol, alongside the previously supported UDP and TCP protocols. This enhancement broadens the applicability of our simulators to a wider range of IoT systems, given the prevalence of MQTT in the IoT sector. The enhancement further enables us to more conclusively evaluate \lang\ by extending our empirical examination to include a new industrial case study focused on a cloud-native IoT-connected vehicle system.

\item~We extend \lang\ to include configurable start times for edge devices by introducing offset variables. This improves our simulators by preventing simultaneous packet transmissions, thereby optimizing the I/O resources and network bandwidth of simulation nodes. Our empirical evaluation demonstrates that this improvement enables our simulators to scale to at least $1.3$ times more IoT devices compared to scenarios where randomness in edge devices' start times is excluded.

\item~We interview two engineers at our industry partner to understand \lang's practical implications. Feedback from these interviews suggests that \lang\ is effective in stress testing IoT cloud systems, saving time, effort, and improving user experience.

\item~We compare simulators derived from \lang\ specifications against two state-of-the-art edge-to-cloud simulators: EMU-IoT~\cite{Ramprasad:19} and Fogbed~\cite{Coutinho:18}, contrasting their designs with ours and highlighting the novel design features of our simulation framework, as shown in Figure~\ref{fig:hierachical}. Our qualitative discussion, combined with empirical results, demonstrates the advantages of our framework in large-scale stress testing.

\item~We present the detailed syntax and semantics of \lang. 
This leads to the specification of four groups of syntactic and semantic checks, which are now supported by the \lang\ editor to assist engineers in developing more accurate \lang\ specifications.

\item~We provide substantial new empirical evidence to demonstrate the effectiveness of simulators generated from \lang\ specifications. We have applied these simulators for stress testing of two real-world systems:  a cloud-based IoT monitoring system developed by our industry partner and an industrial, cloud-native IoT connected vehicle system. Our results show  that these simulators achieve peak performance and can reliably simulate a vast number of edge devices without data loss  when groups of edge devices are containerized through Dockers. Further, we demonstrate how our simulators can help engineers determine in a systematic way  the service capacity of our two  case-study systems. Specifically, we identify  the maximum number of IoT and edge devices that each cloud system can effectively service. To empirically examine whether \lang\ offers added value, we benchmark our simulators against two widely used, open-source baseline tools for stress testing: Apache JMeter~\cite{jmeter} and Locust~\cite{locust}.  Our simulators can simulate 3.5 times more IoT and edge devices compared to JMeter and Locust. This superior performance stems from design features \textbf{F1}, \textbf{F2}, and \textbf{F3} discussed in Section~\ref{subsec:features}.
\end{enumerate}

Our toolset and empirical data are publicly available~\cite{IoTECS} to facilitate reproducibility and so that others can build upon our work.

\subsection{Structure} 
The rest of this article is organized as follows: Section~\ref{sec:usecase} compares the design of our simulation framework with that of two state-of-the-art edge-to-cloud simulators both qualitatively and quantitatively.
Section~\ref{sec:approach} presents our proposed DSL, \lang.  In Section~\ref{sec:eval}, we evaluate the applicability and usefulness of simulators generated from \lang\ specifications for stress testing of two real-world IoT cloud systems. The evaluation assesses different simulator configurations, compares with baseline tools, and synthesizes feedback from practitioners who have used \lang.  Section~\ref{sec:related} discusses related work. Section~\ref{sec:con} concludes the article.

\section{Challenges and Design Decisions}
\label{sec:usecase}
In this section, we evaluate the designs of two IoT simulators from the literature: EMU-IoT~\cite{Ramprasad:19} and Fogbed~\cite{Coutinho:18}. Similar to our work, these simulators focus on edge-to-cloud simulation, making them the closest comparisons available. Below, we detail the design of each of these two simulators and then evaluate the design against the three challenges that respectively motivated the three features discussed in Section~\ref{subsec:features}. For easier reference, we denote the challenges as \textbf{C1} (runtime bloat), \textbf{C2} (bursty communications), and \textbf{C3} (scaling under limited computational resources).

\begin{figure}[t]
\centering
\includegraphics[width=0.65\linewidth]{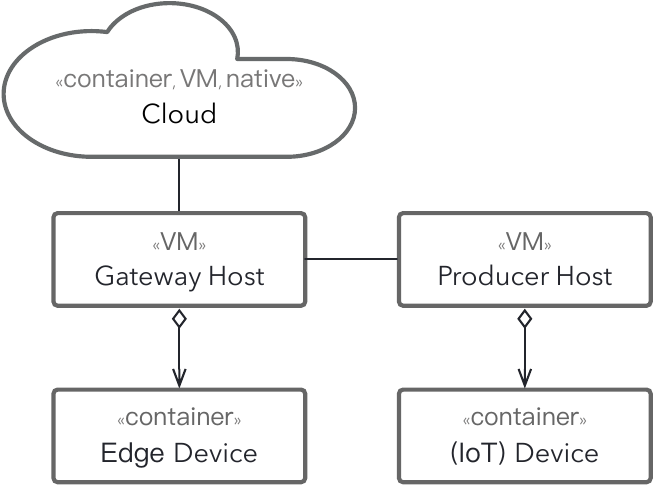}
\caption{The design of EMU-IoT simulator~\cite{Ramprasad:19}}\label{fig:emuiot}
\end{figure}

\textbf{EMU-IoT.} An overview of the design of  EMU-IoT~\cite{Ramprasad:19}  is shown in Figure~\ref{fig:emuiot}. In EMU-IoT, IoT devices are hosted on  virtual machines, known as Producer Hosts. Each IoT device is encapsulated within a Docker container. The IoT devices transmit packets to their corresponding edge devices, which in turn relay the packets to the cloud. The edge devices also function within Docker containers. Gateway Hosts, which operate on virtual machines, host edge devices and enable their connection to cloud systems.

Challenge \textbf{C1}  is not addressed by EMU-IoT, noting that IoT devices operate as standalone processes, and there is no mechanism to prevent excessive system resource consumption when multiple IoT devices are instantiated. As for challenge \textbf{C2}, EMU-IoT employs a load balancer to monitor network resources. Its load balancing policy aims to fully utilize the capacity of hosts by ensuring that the first host is utilized before moving to the next in sequence. This approach, however, does not optimize network usage. Hence,   communication bursts may happen, especially with a large number of simulated IoT and edge devices.

Regarding challenge \textbf{C3}, EMU-IoT isolates \emph{each} IoT and edge device in separate containers, as opposed to our framework, which containerizes \emph{clusters} of edge devices rather than individual ones. The containerization mechanism of EMU-IoT requires significant computational resources for execution, as we demonstrate below.

To further deepen our comparison, we use EMU-IoT to simulate  $10,000$ IoT devices and $100$ edge devices. To capture this system using EMU-IoT, we need to create 10,100 containers. As for the number of virtual machines (VM), we consider three alternative configurations of EMU-IoT: 
\begin{itemize}
\item \textsc{Config1}: $50$ Producer Hosts, each hosting $200$ IoT devices and  two Gateway Hosts, each hosting 50 edge devices; 
\item \textsc{Config2}:  $25$ Producer Hosts, each hosting $400$ IoT devices and two Gateway Hosts, each hosting 50 edge devices; 
\item \textsc{Config3}: $10$ Producer Hosts, each hosting $1000$ IoT devices and two Gateway Hosts, each hosting 50 edge devices.
\end{itemize}

\textsc{Config1}, which is the configuration used to originally evaluate EMU-IoT~\cite{Ramprasad:19}, requires $52$ VMs, \textsc{Config2} requires $27$ VMs, and \textsc{Config3} requires $12$ VMs. We note that hosting these many virtual machines and containers requires powerful server machines. We have executed each configuration of EMU-IoT  so that IoT devices send small size packets ($8$bytes) to their corresponding edge devices, and edge devices forward these packets to the cloud without doing any processing. Table~\ref{tbl:emu-iot} shows the number of packets received by the cloud for each configuration. As shown in the table, zero or very few packets sent by IoT devices are recieved by the cloud, making this tool unsuitable for cloud stress testing.

\begin{table}[t]
  \caption{Simulating $10,000$ IoT devices using the architecture of EMU-IoT.}
  \label{tbl:emu-iot}
  \begin{center}
      \scalebox{0.8}{
      \begin{tabular}{p{4.5cm} p{1cm} p{1cm} p{1cm} p{1cm} }
\toprule
Config&\textsc{Config1}&\textsc{Config2}&\textsc{Config3}\\
\midrule   
\# of packets received by the cloud&845&0&0 \\
percentage of packet loss&97.9\%&100\%&100\%\\
\bottomrule
\end{tabular}}
  \end{center}
\end{table}

\begin{figure}[t]
\centering
\includegraphics[width=0.65\linewidth]{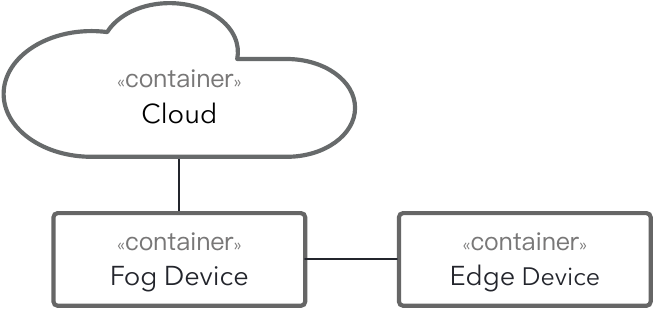}
\caption{The design of the Fogbed simulator~\cite{Coutinho:18}}\label{fig:fogbed}
\end{figure}

\textbf{Fogbed.} The design of Fogbed~\cite{Coutinho:18} is shown in Figure~\ref{fig:fogbed}. Fogbed introduces a fog layer located between the edge and cloud layers.
Like edge devices, fog devices are also used for distributing computation, communication, control and storage closer to end users. This fog layer consists of fog devices that receive data packets from edge devices, perform fog computing, and transmit the processed results to the cloud. In Fogbed, every individual edge device, fog device and cloud node is simulated as a Dockerized node.

Fogbed does not explicitly capture IoT devices. Unlike \lang\ and EMU-IoT, where edge devices transmit data packets from IoT devices to the cloud, Fogbed's edge devices generate data packets as a result of edge computing. Since Fogbed does not deal with IoT devices, the runtime bloat challenge, \textbf{C1}, is irrelevant to Fogbed.

While \lang\ includes design feature \textbf{F2} (Figure~\ref{fig:hierachical}) to optimize network-bandwidth usage, Fogbed does not offer any built-in optimization methods for this purpose. Nevertheless, since Fogbed explicitly captures the network layer, it can be enhanced by network control algorithms that dynamically optimize network bandwidth usage, thus providing low-level facilities to mitigate challenge \textbf{C2}. Fogbed employs containerization at the level of edge devices, fog devices, and cloud nodes. However, it does not cluster edge or IoT devices and lacks means for addressing challenge \textbf{C3}.

\begin{table}[t]
  \caption{Simulating $10,000$ IoT devices using the architecture of Fogbed.}
  \label{tbl:fogbed}
  \begin{center}
      \scalebox{0.8}{
      \begin{tabular}{p{4.5cm} p{1cm} p{1cm} p{1cm} p{1cm}  }
\toprule
Config&\textsc{Config1}&\textsc{Config2}&\textsc{Config3}&\textsc{Config4}\\
\midrule   
\# of packets received by the cloud&1908.2&6386.4&11850&15959.4\\
percentage of packet loss&95.2\%&84.0\%&70.4\%&60.1\% \\
\bottomrule
\end{tabular}}
  \end{center}
\end{table}

Similar to EMU-IoT, we assess Fogbed by simulating 10,000 IoT devices and 100 edge devices. As mentioned earlier, Fogbed cannot directly capture IoT devices. To do so using Fogbed, we extend it with Docker images of edge devices from our simulator. These images symbolically execute the behaviour of IoT devices. Specifically, we use 100 edge device Docker containers, each hosting 100 IoT devices. We consider four different configurations: 
\begin{itemize}
\item \textsc{Config1}, where all edge devices communicate with one fog node; 
\item \textsc{Config2}, where they communicate with two fog nodes; 
\item \textsc{Config3}, where they communicate with five fog nodes; 
\item \textsc{Config4}, where they communicate with ten fog nodes. 
\end{itemize}

The fog nodes are configured to receive packets from edge devices and transmit them to the cloud without any processing. Similar to the previous experiment, we execute Fogbed, where edge devices send 8-byte packets from each IoT device to their corresponding fog nodes, which then forward these packets to the cloud. Table~\ref{tbl:fogbed} shows the number of packets received by the cloud for each configuration. The table indicates that packet loss exceeds 60\% for these configurations, demonstrating that less than half of the packets reach the cloud. This suggests that Fogbed is ineffective for cloud stress testing. As we will demonstrate in Section~\ref{sec:eval}, simulators built using \lang, which leverage features \textbf{F1}, \textbf{F2}, and \textbf{F3} discussed in Section~\ref{subsec:features}, can simulate up to 19,000 IoT devices without packet loss, using setups similar to those in the experiments of this section in terms of packet size and sending frequency.

Based on our analysis in this section, we conclude that existing IoT edge-to-cloud simulators have not been developed with stress testing as one of their intended use cases. While the existing simulators may suffice for small-scale testing involving tens or a few hundred devices with updates happening in the order of minutes, they are inadequate for high-throughput IoT systems that handle thousands of devices and require frequent updates in the order of seconds or sub-seconds.
For this reason, we empirically compare the performance of simulators generated from \lang\ specifications with two widely used stress testing tools, JMeter and Locust, rather than IoT edge-to-cloud simulators in Section~\ref{sec:eval}.

\section{Simulation Framework}
\label{sec:approach}
In this section, we introduce our simulation framework. The framework is centred around a domain-specific language (DSL) named \lang\ (IoT Edge-to-Cloud Simulation Language). Specifications written in \lang\ are translatable into Java, resulting in fully operational simulator code once compiled. This generated simulator can then be executed to produce simulation results. \lang\ is built on top of the conceptual model shown in Figure~\ref{fig:metamodel}. This conceptual model aims to support a scalable and parameterizable architecture for capturing communication between edge devices and cloud applications, enabling simulation-based stress testing of edge-to-cloud solutions. We present our conceptual model in Section~\ref{subsec:metamodel}, where we further elaborate on our decisions regarding the behaviour of our simulation framework related to the model's concepts. In Section~\ref{subsec:dsl}, we discuss \lang's syntax and usage, and in Section~\ref{subsec:check} -- the sanity checks we have implemented in the language to ensure the consistency of \lang\ specifications.

\subsection{The \lang\ Conceptual Model}
\label{subsec:metamodel}

\begin{figure*}[t]
    \begin{center}
    \includegraphics[width=.65\linewidth]{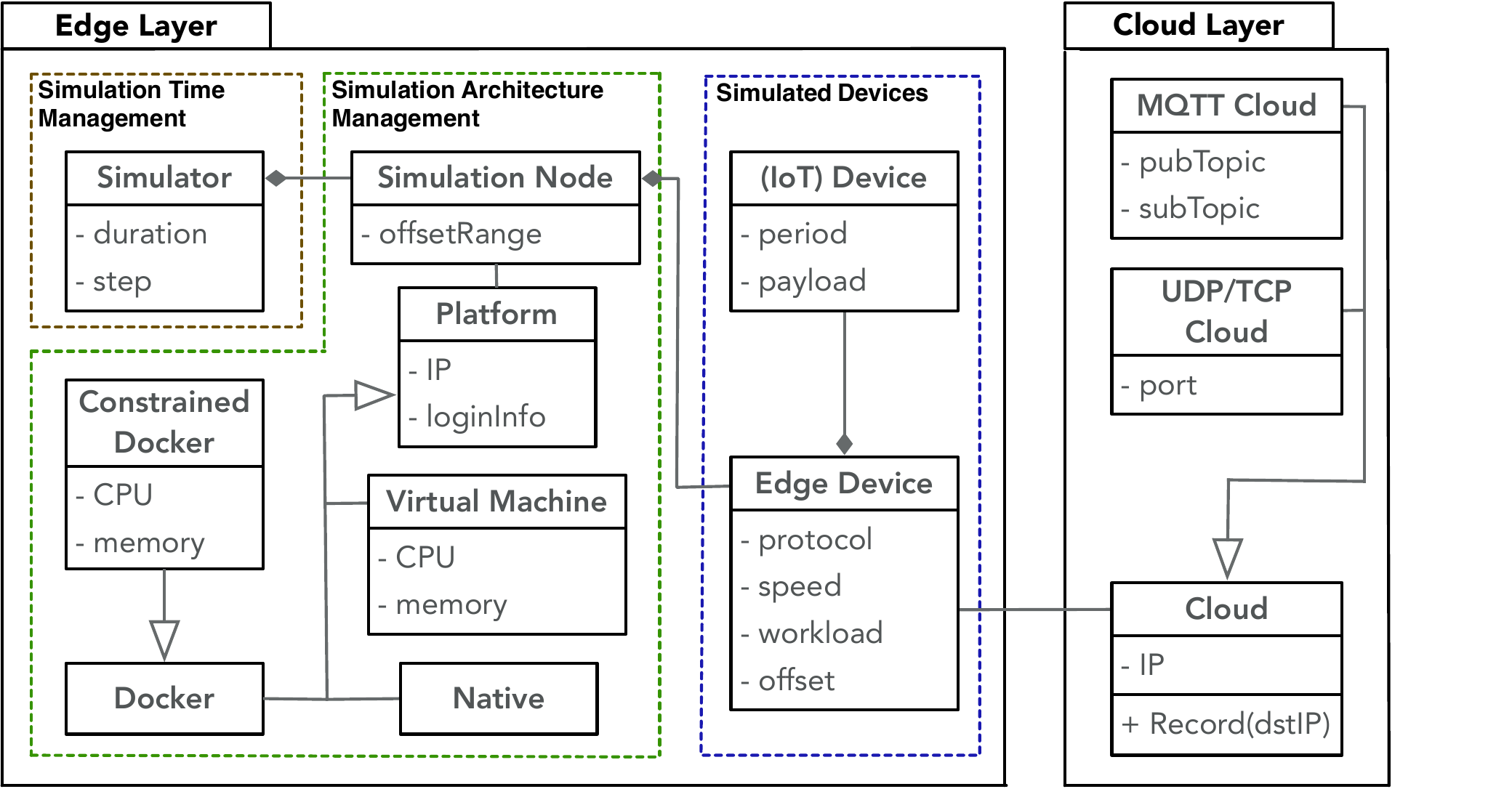}
    \end{center}
    \caption{\lang's underlying conceptual model.}
    \vspace*{-.3cm}
    \label{fig:metamodel}
\end{figure*}

The concepts in the conceptual model of  Figure~\ref{fig:metamodel} are arranged  under two packages: Edge Layer and Cloud Layer. These two layers communicate through a network in a real IoT system. To ensure that the network does not act as a limiter for how far simulated edge devices can stress the cloud systems, we assume that the edge and cloud layers are connected via  an ideal network with minimally low transmission time and no loss. 
In our conceptual model,  the topology of this network is abstracted away, and the network-related information is limited to IPs and communication protocols. In addition, in the case of UDP and TCP protocols, one needs to specify the port numbers, and in the case of MQTT -- the publish topics and the subscribe topics.  This information is captured as necessary within the edge and cloud concepts.

\vspace*{0.3em}\textbf{Cloud Layer}. The \texttth{Cloud} concept shown in  Figure~\ref{fig:metamodel} represents a cloud application under test. In \lang, \texttth{Cloud} is specialized into \texttth{MQTT Cloud} and \texttth{UDP/TCP Cloud}. For an edge-to-cloud simulator to be able to connect to the UDP/TCP cloud under test, we need the \texttth{IP} address of the cloud's host machine and the \texttth{port} at which the cloud receives incoming data. For an MQTT cloud, we need the \texttth{IP} address of the MQTT broker, the publish topic(s) and the subscribe topic(s) for sending packets to the cloud and receiving packets from the cloud, respectively.  The \texttth{Record(dstIP)} method is responsible for collecting incoming and outgoing packets in the cloud. By default,  \texttth{dstIP} is set to the IP address of the cloud's host machine when working with a UDP/TCP cloud, to that of the broker when working with an MQTT cloud. The default \texttth{dstIP} value enables the \texttth{Record(dstIP)} method to collect incoming and outgoing packets on the cloud's host machine for UDP/TCP and on the broker for MQTT. 
For industrial systems,  the cloud may include several components, each deployed on a separate container~\cite{Pahl:19}. In order to assess the performance of a specific cloud component along the path leading to the container of that component, we need to assign the IP address of that container to the \texttth{dstIP} variable of the \texttth{Record(dstIP)} method. For example, MQTT clouds often include a broker and several microservices deployed on separate containers~\cite{Dizdarevic:19,Aksakalli:21}. Our industrial case study  is an example of such cloud architecture. We will discuss in Section~\ref{subsec:expdesign} how we set \texttth{dstIP} to ensure that we asses the performance of the desired cloud component during stress testing. In \lang, we can test several cloud applications simultaneously by instantiating \texttth{Cloud} multiple times. 

\textbf{Simulation Time Management.} The top-level \texttth{Simulator} concept in Figure~\ref{fig:metamodel} captures the attributes required for managing time. We execute each simulation for a given duration and divide this duration into equal time steps. We use the \texttth{duration} and \texttth{step} attributes to refer to the total simulation duration and to the duration of individual time steps, respectively.

\begin{figure*}
    \centering\includegraphics[width=0.72\linewidth]{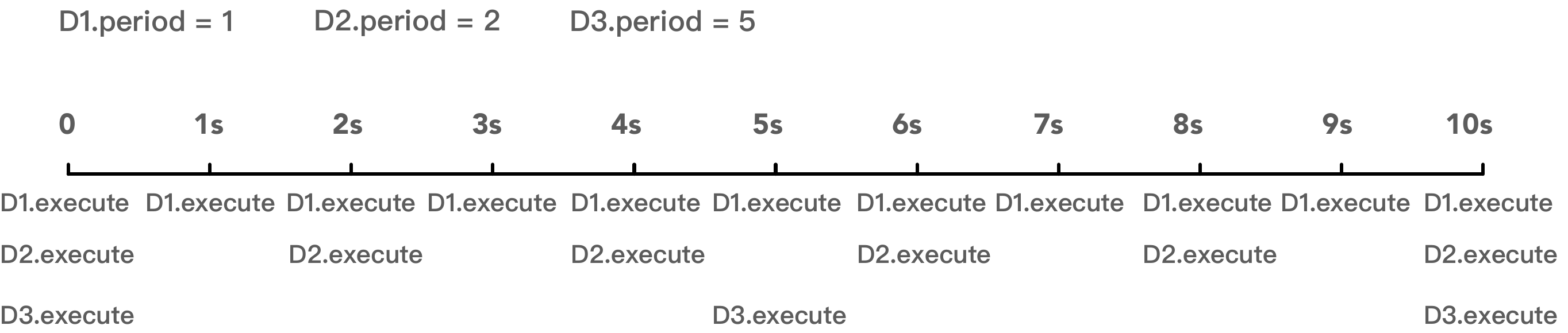}
    \caption{An example simulation run representing the execution of three IoT devices, D1, D2, and D3, over a simulation duration of $10$s with each step taking $1$s. Each IoT device regularly executes based on its period.}
    \label{fig:example}
    \vspace*{-.3cm}
\end{figure*}

\textbf{Simulated Devices.}  The concepts here are \texttth{IoT Device} and \texttth{Edge Device}. IoT devices encompass sensors (e.g.,  temperature sensors) and actuators (e.g., traffic lights).  IoT devices connect to and communicate periodically with edge devices by wired or wireless links. For example, a temperature sensor may communicate temperature readings to an edge device every $5$s. Edge devices can assume a variety of roles, but their primary function is controlling the data flow at the boundary of a network. Edge devices may perform part of the required processing close to the source (IoT devices) instead of relying on the cloud to do all the processing, which is expensive and can further lead to latency~\cite{Varghese:16}. Edge devices nonetheless usually have limited processing and storage capabilities, and their view on data is limited as well. Edge devices are thus not a substitute for the processing and storage done by the cloud.

It is crucial that  \lang\ should be able to simulate edge devices with high fidelity. For this, we do not need to explicitly simulate IoT devices through individual processes or threads. Neither do we need to distinguish between IoT sensors and actuators, noting that both can have two-way communication with edge devices. Instead, as explained in Section~\ref{subsec:features}, we utilize design feature \textbf{F1}, i.e., the symbolic representation of IoT devices, to address the challenge of run-time bloat. This approach, as we elaborate below, integrates relevant properties of IoT devices into the execution of (simulated) edge devices. For succinctness, where there is no ambiguity, we refer to a sensor or actuator simply as ``device''. In contrast, when we mean an edge device, we use \hbox{the term ``edge device''.}

Each instance of \texttth{IoT Device} has a \texttth{period} attribute, indicating the time interval (in terms of the number of simulation steps) between two successive executions of the IoT device. For example, Figure~\ref{fig:example} shows a simulation run where the  duration is $10$s (\texttth{duration} = 10) and each step takes $1$s (\texttth{step} = 1s),  dividing the total duration into ten sequential time slots. An IoT device with \texttth{period} $= 1$ executes its function every step and the one with \texttth{period} $= 2$ executes every other step.
IoT devices, whenever they execute, generate a data packet whose size is specified by the \texttth{payload} attribute.

Each instance of \texttt{Edge Device} owns a group of IoT devices. At every time step, each edge device collects the data packets generated by its associated IoT devices. As mentioned earlier, in  \lang, the IoT device concept does not induce executable entities. Instead, each instance of an edge device, which is an executable entity in \lang, is responsible for emulating the behaviour of the IoT devices associated with it. We provide a precise characterization of the behaviour of edge devices later, when \hbox{we discuss Algorithm~\ref{algo:edgedevice}.}

Edge devices send the collected packets to the cloud following the communication protocol indicated by the \texttth{protocol} attribute in \texttt{Edge Device}. 
In parallel with sending packets, the edge device performs two more operations: (a)~receiving data from the cloud, and (b)~performing compute-intensive operations (i.e., edge computing). Edge devices use a \texttth{workload} attribute to specify the time duration, measured in milliseconds, seconds or minutes, spent on compute-intensive operations.

Even though edge devices operate at every time step, in real-world applications, they do not all start their operation at the same instant,  and may send data packets with some time gaps within a time step~\cite{Mani:18}. For every edge device, we use two attributes \texttth{offset} and \texttth{speed}  to capture the randomness in start time and the time gap between the sending of packets, respectively. For instance, if we set \texttth{offset} of an edge device to 10ms, at every time step, the edge device waits for 10ms before it starts its operation. In this paper, we assign offset values drawn from a uniform distribution to simulate scenarios where edge devices executes with minor and random time delays relative to each other. The  \texttth{speed} attribute specifies how many packets an edge device sends to the cloud per simulation time step. To illustrate, let the simulation time step be $1$s. If we set \texttth{speed} to $100$, the edge device sends \emph{up to} $100$ packets over $1$s. That is, the edge device waits for $10$ms after sending each packet to the cloud. In the above, we say ``up to'' because the edge device may send fewer packets if its associated IoT devices produce less packets than what the speed attribute allows to be transmitted. Similarly, if we set \texttth{speed} to $1000$, the edge device waits for $1$ms after sending each packet. To indicate that an edge device can send packets at maximum speed over the time step duration (i.e., without any waiting after sending a packet),  one can set \texttth{speed} to a designated value, \texttt{MAX}. When an edge device's \texttt{speed} value is set to ``MAX'', the edge device consecutively sends all of its packets without any delays between each transmission.

If a large number of (simulated) edge devices attempt to send all their packets to the cloud without any time gap and any offsets, i.e., offsets set to zero, the simulator may not be able to cope with all the send requests, or these requests may congest the network. In both situations, one may experience packet loss or latency at the (simulated) edge or within the network. In view of our goal, which is to stress test cloud systems, we need to make sure that we do not incur packet loss or latency at the edge or in the network.  Using the speed attribute, we can pace the send requests in a way that would prevent the simulator and the network from being overwhelmed. By offsetting the start time of edge devices, we avoid situations  where all edge devices within the same simulation node send packets to the cloud simultaneously. Further, by staggering the start times of edge devices, we reduce the  utilization of I/O resources of the simulation node and the network. These two attributes, \texttth{offset} and \texttth{speed}, belong to design feature \textbf{F2}. As explained in Section~\ref{subsec:features}, this feature helps mitigate bursts in edge-to-cloud data transmission, allowing
us to rule out the simulator and the network as the root cause for packet loss and thus localize packet-loss occurrences to the cloud. In our empirical evaluation of Section~\ref{sec:eval}, we demonstrate the impact of \texttt{speed} and \texttth{offset} on our results.

\begin{algorithm}[t]
   \footnotesize
\DontPrintSemicolon
\SetSideCommentRight
\SetNoFillComment
    \SetKwBlock{DoParallel}{run in parallel:}{end}
    \SetKwBlock{sendp}{Send Process:}{end}
    \SetKwBlock{receivep}{Receive Process:}{end}
    \SetKwBlock{computep}{Compute Process:}{end}
    \KwIn{\;
    $T$: Simulation  duration\;\\
    $q$: Size of the simulation time step\;\\
    $n$: Number of IoT devices associated to $e$\; \\
    }
    sleep($\mathit{e.offset}$)\\
    \DoParallel{
       \sendp{
       \For{$i = 0$ to $\frac{T}{q}$}{
         startTime $\gets$ getWallClockTime() \;
         \For{every $j \in \{1, \ldots, n\}$}{
        
        \If { ($i$ \mbox{mod} e.devices[j].period == 0)} 
        {Send $\mathit{e.devices[j].payload}$ to $\mathit{e.cloud}$ using $\mathit{e.protocol}$\;
        \If {$\mbox{e.speed}$ != MAX } {
        sleep($\frac{\mathit{q}}{\mathit{e.speed}}$) }}
        currentTime $\gets$ getWallClockTime() \; 
        \If {(q $\leq$ currentTime - startTime)} {
        Break out of the loop (of line~6) \Comment{go to the next step}\;}}
        }} 
        \receivep {
        Receive packets from $\mathit{e.cloud}$ \;
        } 
        \computep{
        Perform CPU-intensive operations for duration $\mathit{e.workload}$ \Comment{do edge computing}\; 
        }
    }
    \caption{The behaviour of Edge Device $e$.}
    \label{algo:edgedevice}
\end{algorithm}

Algorithm~\ref{algo:edgedevice} formally specifies the behaviour of edge devices in \lang. An edge device, denoted by `e' , waits for a duration determined by $e$'s offset parameter (line~1). Then, the 
edge device performs three operations in parallel: sending packets, receiving packets and performing (edge) computing. At every time step $i$ between $0$ to $\frac{T}{q}$ where $T$ is the simulation duration and $q$ is the size of the step, the edge device $e$ checks each of its associated IoT devices (line~6). If the period of an IoT device indicates that it should be executed at step $i$ (line~7), then $e$ sends the associated payload to the cloud (line~8). After sending the data, $e$ may wait for some time gap whose size depends on the speed parameter of $e$ (lines~9-10). The loop for sending the IoT devices' data to the cloud (lines~6-13) ends as soon as the time step $q$ has elapsed (line~12-13). As a result, if the time step is too short or too many IoT devices are associated to $e$, some may fail to send their data packets to the cloud.

The receive method is used for receiving packets transmitted from the cloud to edge device $e$ (lines 14-15). To simulate edge computing in \lang, edge devices perform some compute-intensive operation for the duration specified by the \texttth{workload} attribute (lines 16-17). We note that, to properly simulate edge computing, one needs to explicitly perform operations that keep the edge CPU busy so that the CPU is not allocated to other parallel tasks. To this end, we perform a series of floating-point operations. 
Algorithm~\ref{algo:edgedevice} stops when the simulation duration $T$ elapses.

\textbf{Simulation Architecture Management.} An important factor in edge-to-cloud simulation is the ability to simulate a large number of IoT and edge devices. As discussed above, each edge device manages the sending of the packets related to its IoT devices sequentially and within the simulation time step $q$. This assumption matches practice, since in the real-world, edge devices directly handle their related IoT devices. The edge devices themselves, however, run in parallel and are independent from one another. The concepts under Simulation Architecture Management (Figure~\ref{fig:metamodel}) enable a more orderly handling of concurrency, in turn allowing us to optimize the number of edge devices that can run in parallel.

To maximize the capacity of our simulator in terms of the number of parallel edge devices under limited computational resources, we introduce an additional tier into our simulator's architecture to group the edge devices into clusters. This tier, corresponding to design feature \textbf{F3} in Section~\ref{subsec:features}, is represented by the \texttth{Simulation Node} concept in the model of Figure~\ref{fig:metamodel}. The \texttth{Simulation Node} concept includes the offsetRange attribute  which limits the offset values of its associated edge devices.
That is, the offsetRange value indicates the maximum offset values that edge devices associated to the simulation node can have. We specify offsetRange as a percentage of the simulation time step. To illustrate, let the simulation time step be $1$s. If we set \texttth{offsetRange} to $20$\%, the offset value for all edge devices associated to this simulation node is a value between $0$ to $0.2$s, or $0$ to $200$ms.

Each simulation node is  associated with a platform, represented by the \texttt{Platform} concept. Edge devices that belong to a given simulation node all execute in parallel and on the platform  of the simulation node. The \texttt{Platform} concept has as attributes \texttth{IP} 
and \texttth{loginInfo} to enable access to the platform host machine. 
In \lang, \texttt{Platform} is  specialized into \texttt{Native} (i.e., running on a native operating system), \texttth{Virtual Machine} and \texttth{Docker}. For a virtual machine (VM), we need to specify the \texttth{CPU} and \texttth{memory} attributes, respectively indicating the CPU and memory allocated  to the VM.  
Docker containers can be unconstrained or constrained. For the latter (i.e., \texttth{Constrained Docker}), similar to a VM, we need to specify the \texttth{CPU} and \texttth{memory} attributes.  

The idea of clustering edge devices into  simulation nodes in order to scale simulators is inspired by the notion of ``super indivduals'' in massively multi-agent simulators~\cite{Scheffer:95}. 
In our context, such clustering provides a mechanism to reduce contention over CPU, memory and ports. For example, instead of running 100 parallel edge devices on a single computer, we create ten Dockers by dividing the single computer resources (CPU, memory and ports) between the Dockers equally. We then use each Docker to host ten (simulated) edge devices. As we will demonstrate in our evaluation of Section~\ref{sec:eval}, the Docker option considerably improves the scalability of our simulator, compared to the native option. 

Another motivation for defining simulation nodes is enabling the restriction of simulation resources. By setting limits on CPU and memory for virtual machines and (constrained) Dockers, users can more effectively manage simulation costs, especially when utilizing third-party computation resources. We will revisit this topic in our conclusion section (Section~\ref{sec:con}), where we discuss cost-awareness for simulation.

\begin{algorithm}[t]
\footnotesize
\DontPrintSemicolon
\SetSideCommentRight
\SetNoFillComment
\SetKwBlock{DoParallel}{run in parallel:}{end}
    \KwIn{\;
    $m$ : Number of edge devices associated to $\mathit{sn}$\;
    $q$ : Size of the simulation time step\;\\
    }
    \For {every $i \in \{1, \ldots, m\}$} 
    {initialize $\mathit{sn}.edgeDevice[i]$ \;   
    select a random number $ran$ in [0 .. $\mathit{sn}.\mathit{offsetRange}\cdot q$] \;
    $\mathit{sn}.edgeDevice[i].\mathit{offset}$ $\gets$ $ran$
    \;}
    
    {In parallel, run all $\mathit{sn}.edgeDevice[i]$ ($i \in \{1, \ldots, m\}$) \Comment{run Algorithm~\ref{algo:edgedevice} for each $\mathit{sn}.edgeAgent[i]$} }
    \caption{The behaviour of Simulation Node $\mathit{sn}$.}
    \label{algo:simnode}
\end{algorithm}

The behaviour of \texttth{Simulation Node} is specified in Algorithm~\ref{algo:simnode}. Each simulation node, upon creation, initializes each of its associated edge devices (lines 1-2). 
To assign the offset parameter for each edge device associated with the simulation node (line 3-4), we generate a random number within the following range: \hbox{$[0 .. \mathit{offsetRange} \cdot q]$}, where $q$ is the simulation time step. 
To ensure that all values within this range are equally likely to be selected, we use a discrete uniform distribution for generating (integer) offset values. Subsequently, 
the edge devices are executed in parallel (line 5). That is, Algorithm~\ref{algo:edgedevice} is called for each edge device. 

Algorithm~\ref{algo:simulator} shows the overall behaviour of our simulator. The simulator starts by initializing the platforms of the simulation nodes (line~3). It then transfers the simulation code to the platforms (line~4). Next, the simulation nodes start to run in parallel on their respective platforms (line~5). The nodes run until the simulation duration elapses (line~6), at which point the nodes and their platforms are cleaned up (lines~7-9).

Our simulator can be configured to represent different numbers of IoT and edge devices. As these numbers increase, the application(s) in the cloud layer are put under more stress. In this way, our simulator makes it possible to determine how far the cloud application(s) under test can be stretched without degradation in their quality of service.
Upon the termination of a simulation round, we collect certain metrics, defined in Section~\ref{subsec:metric}. These metrics help determine (1)~whether, given the computational resources  available, the simulator can mimic the desired numbers of IoT and edge devices without getting overwhelmed, and (2)~whether the cloud system(s) are able to process and respond in reasonable time to all the messages received from the simulator.

\begin{algorithm}[!t]
\footnotesize
\DontPrintSemicolon
\SetSideCommentRight
\SetNoFillComment 
\SetKwBlock{DoParallel}{run in parallel:}{end}
 \KwIn{\;
    \emph{l}: Number of simulation nodes\;\\
    }
    
    \For {every $i \in \{1, \ldots, l\}$} 
   {  Let $p$ denote $\mathit{sim}.\mathit{sn}[i].\mathit{platform}$ \;
      Initialize $p$ \;
      Transfer $\mathit{sim}.\mathit{sn}[i]$ to $p$ using $\mathit{p.IP}$ and $\mathit{p.loginInfo}$ \;}

    In parallel, run all $\mathit{sim}.\mathit{sn}[i]$ ($i \in \{1, \ldots, l\}$) \Comment{run Algorithm~\ref{algo:simnode} for each $\mathit{sim}.\mathit{sn}[i]$} \;
    sleep ($\mathit{sim.duration}$) \;
    
    \For {every $i \in \{1, \ldots, l\}$} 
    {
    Terminate $\mathit{sim}.\mathit{sn}[i]$ if it is still running  \;
    Clean up $\mathit{sim}.\mathit{sn}[i].\mathit{platform}$ and $\mathit{sim}.\mathit{sn}[i]$}
    \caption{The bahviour of Simulator $\mathit{sim}$.}
    \label{algo:simulator}
\end{algorithm}

\subsection{The \lang\ DSL}
\label{subsec:dsl}
\lang\ aims to support  practitioners in creating simulators that are instances of  the conceptual model of Figure~\ref{fig:metamodel}. 
We provide an implementation of \lang\ using Xtext~\cite{Xtext}.
Specifically, we use Xtext's grammar language to define \lang's grammar.
To generate executable simulators from specifications written in \lang, we use Xtend~\cite{Xtend}. More precisely, we use Xtend to retrieve all the objects defined in an \lang\ specification; these objects in turn drive the instantiation of a number of a-priori-defined Xtend templates. Our Xtend templates include: (1) Java-based implementations of the algorithms in Section~\ref{subsec:metamodel}, (2) scripts for setting up / starting virtual machines and containers, (3) scripts for uploading and running simulator code on remote platforms, (4) scripts for downloading simulation results from simulation nodes, and (5)~scripts for analyzing and reporting simulation results. 

In the rest of this section,  we provide the syntax of \lang\ using the Backus-Naur Form (BNF) notation: non-terminal symbols are surrounded by angle brackets, while terminal symbols are enclosed in single quotes; ::= means that the symbol on the left must be replaced with the expression on the right; a vertical bar is used to separate alternatives; a star stands for zero or more occurrences; a plus indicates  one or more occurrences; square brackets represent optional elements; parentheses are used for grouping; literals are enclosed in `'.
Below, we present and exemplify the syntax of \lang. Full details about the design of \lang\ can be found in our  artifacts~\cite{IoTECS}.

\textbf{Cloud.}
The syntax of \texttth{Cloud} is defined as follows:

 \begin{grammar}
<Cloud> ::=  `Cloud:' <ID> `{'  \hfill ($\alpha$1) \\
 `IP:' <IP> \hfill ($\alpha$2) \\
 ( `port:'  <integer> | \hfill ($\alpha$3) \\
 `pubTopic:' <string>  $[$`, subTopic:' <string>$]$)  \hfill ($\alpha$4) \\
 $[$ `Methods: {' `Record(' <IP> `)' `}' $]$   \hfill ($\alpha$5)\\
 `}'\par
<ID> ::= (`a'-`z' | `A'-`Z' | `0'-`9')+ \hfill ($\alpha$6)\par
<IP> ::= (`0'-`9')+`.'(`0'-`9')+`.'(`0'-`9')+`.'(`0'-`9')+ \hfill ($\alpha$7) 
 \end{grammar}

 The above syntax requires a unique \texttt{ID} to be assigned to each cloud instance (line~$\alpha$1). \lang\ supports two types of IoT cloud systems: one uses MQTT as the messaging protocol and the other uses UDP or TCP. For the first type of cloud systems, \lang\ requires the \texttt{IP} address of the MQTT broker, a topic for publishing messages to the broker and a topic for subscribing responses from the broker (lines $\alpha$2 - $\alpha$4). The subscribe topic is optional because some cloud systems may not send packets back to edge devices.  For the TCP or UDP cloud systems, the \texttt{IP} address and a \texttt{port} of the cloud's host machine are needed to enable communication between the edge devices and the cloud (lines $\alpha$2 - $\alpha$3). 
 The optional \texttt{`Record($\langle$IP$\rangle$)'} method (line~$\alpha$5) initiates a network-traffic monitor for the specified \texttt{IP}. This enables the recording of all incoming and outgoing packets associated with \texttt{IP}. The recorded data can  be used for evaluating stress testing metrics like packet loss and delay.
 
Figure~\ref{fig:dslCloudExample} shows a snippet of an \lang\ specification specifying two \texttth{Cloud} instances. For cloud instance C1, the \texttth{IP} and \texttth{port} attributes respectively specify the IP address of the cloud's host machine and the port at which incoming UDP or TCP packets are received. In contrast, C2 is an instance of MQTT cloud. The \texttth{pubTopic} and \texttth{subTopic} attributes specify the publish topic and the subscribe topic of the MQTT broker (within the cloud) respectively.  We note that the code for instances of \texttth{Cloud} is not meant to be generated by \lang. Rather, each \texttth{Cloud} instance represents an \emph{existing} cloud system that needs to be stress-tested.

\begin{figure}[t]
\centering
\begin{lstlisting}[basicstyle=\scriptsize,style=Spec]
Cloud:C1 {
	IP:192.168.0.2
	port:1883
} 
Cloud:C2 {
	IP:192.168.0.3
	pubTopic:pub
  subTopic:sub
}
\end{lstlisting}
\caption{Example cloud systems defined using \lang.}
\label{fig:dslCloudExample}
\end{figure}

\textbf{Simulator.}
Each \lang\ specification has exactly one instance of the \texttt{Simulator} concept.
The syntax of \texttth{Simulator} is defined as follows:

\begin{grammar}
<Simulator> ::= `Simulator:' `{' \hfill ($\beta$1)\\
`duration:' <integer> <TimeUnit> \hfill ($\beta$2)\\
`step:'  <integer> <TimeUnit> \hfill ($\beta$3)\\
`simulationNodes:{'<ID>`['<integer>`]'\hfill ($\beta$4)\\
\rule{2.7cm}{0pt} (`,'<ID>`['<integer>`]')* `}' \\
`}'\par
<TimeUnit> ::= `ms' | `s' | `m' | `h' \hfill ($\beta$5) 
\end{grammar}

In the above syntax, \texttth{duration} (line $\beta$2) is used to specify the total duration of simulation.  On the other hand, \texttth{step} (line $\beta$3) is used to declare the simulation time step. 
To illustrate, consider the snippet in Figure~\ref{fig:dslSimulatorExample}. 
Here, the simulator has its \texttt{duration} set to 10s and its \texttt{step} to 1s; this results in dividing the simulation into 10 steps of equal length with each step running for 1s. The time unit for \texttt{duration} and \texttt{step} can be in milliseconds (ms), seconds (s), minutes (m) or hours (h) (line $\beta$5). A simulator needs to declare its simulation nodes
which are expressed as a comma-separated list after the keyword \texttt{`simulationNodes'} (line $\beta$4). Each item in the list is the ID of a simulation node type followed by the number of node instances of this type in square brackets. (see line~4 of Figure~\ref{fig:dslSimulatorExample} for an example). On this line, we are stating that the simulator has six simulation nodes: five nodes of type \texttt{SN1} and one node of type \texttt{SN2}. We next illustrate how to define the simulation node types (in this case, \texttt{SN1} and \texttt{SN2}) and their associated execution platforms.

\begin{figure}[t]
\begin{lstlisting}[style=Spec]
Simulator: {
	duration:10s
	step:1s
	simulationNodes:{SN1[5],SN2[1]}
} 
\end{lstlisting}
\caption{Example simulator defined using \lang.}
\label{fig:dslSimulatorExample}
\end{figure}

\textbf{Simulation nodes and platforms.} As discussed in Section~\ref{subsec:metamodel}, \lang\ uses the notion of simulation node for grouping simulated edge devices and to further specify the platform on which a group of edge devices runs. 
The syntax of \texttth{Simulation Node} is defined as follows:

\begin{grammar}
<SimulationNode> ::= `SimulationNode:' <ID> `{' \hfill ($\gamma$1)\\
`platform:' <ID>  \hfill ($\gamma$2)\\
`offsetRange:' <integer>`\%' \hfill ($\gamma$3)\\
`EdgeDevices:{'<ID>`['<integer>`]' \hfill ($\gamma$4)\\
\rule{2cm}{0pt} (`,'<ID>`['<integer>`]')* `}'\\
`}'
\end{grammar}

Each instance of a simulation node must have a (unique) ID  (line $\gamma$1). The ID of the platform on which the simulation node runs is indicated after the keyword `platform' (line $\gamma$2). The keyword `offsetRange'  specifies an upper bound for the offset values for all the  edge devices related to this simulation node (line $\gamma$3). Recall from Section~\ref{subsec:metamodel} that, by introducing offsets, one can circumvent bursts, i.e., situations where all edge devices transmit at the same time. Using offsets increases the realism of simulations for applications where bursts are unlikely to occur or are explicitly mitigated. 
In the snippet of Figure~\ref{fig:dslHierarchyExample}, we define two simulation node types, \texttt{SN1} and \texttt{SN2}, each having a platform, an offsetRange and  a set of edge devices. Similar to our convention for specifying the simulation nodes contained in a simulator (see Figure~\ref{fig:dslSimulatorExample}), we define the edge devices contained in a simulation node via an edge-device type and a number of instances (line $\gamma$4).  For example, in Figure~\ref{fig:dslHierarchyExample}, \texttth{SN1} has ten edge devices: seven of type \texttth{E1} and three of type \texttth{E2} (we will momentarily illustrate the specification of edge-device types). For \texttth{SN1}, the \texttth{offsetRange}  is set to $20$\% of the time step duration, i.e., 200ms, while for  \texttth{SN2}, \texttth{offsetRange} is set to $60$\% of the time step duration, i.e., 600ms. Hence, the offsets of edge devices belonging to \texttth{SN1} are selected from the range $[0..200ms]$, while the offsets of edge devices belonging to \texttth{SN2} are selected from the range $[0..600ms]$.

\begin{figure}[t]
\begin{lstlisting}[style=Spec]
SimulationNode: SN1 {
	platform:P1
  offsetRange:20%
	EdgeDevices:{E1[7],E2[3]}
} 
SimulationNode: SN2 {
	platform:P2
  offsetRange:60%
	EdgeDevices:{E1[30]}
} 

Platform: P1{
	type: Docker
	IP: 192.168.0.4
	username: user2
	CPU: 4
	memory: 2G
}

Platform: P2{
	type: Docker
  CPU: 2
  memory: 2G
}
\end{lstlisting}
\caption{Example simulation nodes and platforms defined using \lang.}
\label{fig:dslHierarchyExample}
\end{figure}

A platform represents either a physical or virtual host for a simulation node. The syntax of \texttth{Platform} is defined as follows:

\begin{grammar}
<Platform> ::= `Platform:' <ID> `{' \hfill ($\delta$1)\\
`type:' <Type>  \hfill ($\delta$2)\\
$[$`IP: ' <IP> \rule{0.4cm}{0pt}`userName:' <string> $]$ \hfill ($\delta$3)\\
$[$`CPU:' <integer> \rule{0.4cm}{0pt}
`memory:' <integer> `G' $]$ \hfill ($\delta$4)\\
`}' \par
<Type> ::= `Native' | `Docker' | `VM' \hfill ($\delta$5) 
\end{grammar}

Each platform has a type (line $\delta$2) that assumes one of the following  values: \texttth{Native}, \texttth{VM} or \texttth{Docker} (line $\delta$5). To be able to run a simulation node when its platform is remote (i.e., not the local host), the simulation code and scripts need to be transferred to and set up on the platform first. For this purpose, we need to specify an IP (line $\delta$3) and login credentials (username and password) in instances of the \texttth{Platform} concept.
As a security measure, we elect to not capture passwords in \lang; this is to avoid storing sensitive information in plain text.  Instead, users are prompted directly for a password by any host involved in the simulation process, as per the scripts generated from an \lang\ specification. The \texttth{Platform} concept is illustrated in the snippet of Figure~\ref{fig:dslHierarchyExample}. Here, \texttth{P1}, which is of type \texttth{Docker}, is specified on lines 12-18. When a platform is hosted locally, one does not need \texttt{IP} and \texttt{loginInfo}. This is illustrated by \texttt{P2} -- another platform of type \texttth{Docker} -- on lines 20-24 of the snippet. 
For a \texttth{Docker} platform that is constrained and for any \texttth{VM} platform, we specify the required attributes indicated in our conceptual model of Figure~\ref{fig:metamodel}. For example, platform \texttth{P1} is a constrained Docker with \hbox{4 CPUs and 2G of memory.}

\textbf{Edge and IoT devices.}
The syntax for defining an \texttth{Edge Device} is as follows:

\begin{grammar}
<EdgeDevice> ::= `EdgeDevice:' <ID> `{' \hfill ($\epsilon$1)\\
`protocol:' <Protocol>  \hfill ($\epsilon$2)\\
`speed:' <integer> | `MAX'\hfill ($\epsilon$3)\\
`cloud:' <ID> \hfill ($\epsilon$4)\\
`devices:{'<ID>`['<integer>`]' \hfill ($\epsilon$5)\\
\rule{1.4cm}{0pt} (`,'<ID>`['<integer>`]')* `}'\\
`workload:' <integer><WorkloadTimeUnit> \hfill ($\epsilon$6) \\
`}'\par
<Protocol> ::= `UDP' | `TCP' | `MQTT' \hfill ($\epsilon$7) \par
<WorkloadTimeUnit> ::= `ms' | `s' | `m' \hfill ($\epsilon$8) \par
\end{grammar}

Each edge device has a set of IoT devices  (line $\epsilon$5) associated to it. For example, in the snippet of Figure~\ref{fig:dslEdgeDeviceExample}, edge-device type \texttth{E2} is associated with 10 IoT devices of type \texttth{D1} and 20 IoT devices of type \texttth{D2} (we will shortly exemplify IoT-device types).

Each edge device communicates with one cloud system which is indicated after the keyword `cloud' (line $\epsilon$4). For example, in Figure~\ref{fig:dslEdgeDeviceExample}, \texttth{E1} is specified as communicating with cloud system \texttth{C1}. The protocol used by an edge device for communication with the cloud is captured by the \texttth{protocol} attribute (line $\epsilon$2). Currently, \lang\ supports the \texttth{UDP}, \texttth{TCP} and \texttth{MQTT} protocols. The \texttth{speed} attribute  describes the number of packets sent to the associated cloud system in one time step. If one does not want to constrain the number of packets, 
the \texttth{speed} attribute should be set to the designated value of `MAX'. For instance, in Figure~\ref{fig:dslEdgeDeviceExample}, the \texttth{speed} is set to 100 for \texttth{E1} and 1000 for \texttth{E2}. Since \texttt{step} has been set to 1s (see Figure~\ref{fig:dslSimulatorExample}), \texttth{E1} waits 10ms between sending two consecutive packets; this wait time is 1ms for \texttth{E2}. The \texttth{workload} attribute (line $\epsilon$6) indicates the amount of edge computing to be done by the edge device (lines 16-17 of Algorithm~\ref{algo:edgedevice}). 
The time unit for \texttt{workload} can be milliseconds (ms), seconds (s) or minutes (m). Note that, in parallel with sending packets and  edge computing, an edge device also receives packets from the cloud (lines 14-15 of Algorithm~\ref{algo:edgedevice}). The protocol for receiving data from the cloud is the same as that used for sending data to the cloud (i.e., is as specified by the \texttth{protocol} attribute).

\begin{figure}[t]
\begin{lstlisting}[style=Spec]
EdgeDevice: E1 {
	protocol:TCP
	speed:100
	cloud:C1
	devices:{D1[100]}
} 
EdgeDevice: E2 {
	protocol:MQTT
	speed:1000
	cloud:C2
	devices:{D1[10],D2[20]}
	workload:20ms
} 
\end{lstlisting}
\caption{Example edge devices defined using \lang.}
\label{fig:dslEdgeDeviceExample}
\end{figure}

\texttth{IoT device} is the lowest-level entity in \lang\ and its syntax is defined as follows:

\begin{grammar}
<Device> ::= `Device:' <ID> `{' \hfill ($\zeta$1)\\
`period:' <integer>  \hfill ($\zeta$2)\\
`payload:' (<integer> <PayloadUnit>) | <string>  \hfill ($\zeta$3) \\
`}'\par
<PayloadUnit> ::= `b' | `kb' | `mb' \hfill ($\zeta$4) 
\end{grammar}

Each IoT device (type) has two attributes.
The first one is \texttt{period} (line~$\zeta$2) which determines the frequency of execution. For example, in the snippet shown in Figure~\ref{fig:dslDeviceExample}, device \texttth{D1} generates a packet every second, while device \texttth{D2} generates a packet every two seconds. 
The second attribute is \texttth{payload} (line $\zeta$3). This attribute specifies the actual payload content, e.g., \texttth{payload: "23C"}, or alternatively, the size of the packet that the device sends every time the device executes. In the latter case, the payload unit can be bytes (b), kilo bytes (kb) or mega bytes (mb). When a size unit is indicated for the payload, the content of the payload is generated randomly as per the requested size. This option, illustrated in Figure~\ref{fig:dslDeviceExample}, is convenient when the actual payload is unimportant for simulation purposes (e.g., in a simple cloud storage application).

\begin{figure}[t]
\begin{lstlisting}[style=Spec]
Device: D1 {
	period:1
	payload:60b
} 
Device: D2 {
	period:2
	payload:100b
} 
\end{lstlisting}
\caption{Example (IoT) devices defined using \lang.}
\label{fig:dslDeviceExample}
\end{figure}

\subsection{Sanity Checks}
\label{subsec:check}

This section describes the syntactic and semantic sanity checks implemented in \lang's Eclipse-based text editor. Users get instant errors or warnings for any violations. There are four groups of sanity checks as we explain below:

\textbf{Declaration/usage checking.} 
\lang\ requires explicit declarations for all entities in a specification. For example, to define a \texttt{Simulator} instance, one must specify its simulation node(s), which can be declared before or after the \texttt{Simulator} instance.
Furthermore, \lang\ ensures that every declaration is used at least once in the specification. This prevents unused or redundant entities, which may indicate errors or oversights. Unused declarations trigger warnings.

\textbf{Type checking.} 
\lang\ enforces type checking to ensure consistent and correct usage of entity types throughout a specification. For example, \lang\ expects IDs of simulation nodes to follow the keyword `simulationNodes' when a \texttt{Simulator} instance is being defined. If \lang\ detects IDs of other entity types (e.g., devices or platforms) instead, it reports a type mismatch error.

\textbf{Protocol checking.} \lang\ verifies that the \texttt{Protocol} attribute of an edge device matches the cloud instance type it communicates with. If an edge device has an MQTT cloud instance as its \texttt{cloud} attribute, its communication protocol should be ``MQTT''. Likewise, if it has a UDP/TCP cloud instance as its \texttt{cloud} attribute, its communication protocol should be either ``UDP'' or ``TCP'', depending on the packet type used for cloud communication.

\textbf{Time-step checking.} \lang\ verifies that the simulation time step is a divisor of the simulation duration. This ensures that one has a whole number of steps within the simulation duration.

\section{Evaluation}
\label{sec:eval}
In this section, we evaluate the applicability and usefulness of simulators generated from \lang\ specifications for stress testing IoT cloud systems. We use the term \emph{simulator} to refer to the executable code generated from instantiating the edge-layer concepts of the model of Figure~\ref{fig:metamodel}; we use the term \emph{cloud} to refer to instances of the cloud concept in this model. To evaluate the performance of simulators derived from \lang\ specifications, we use two real-world IoT cloud systems as case studies. The first case study is a cloud-based IoT monitoring  system developed by one of our industry partners; this cloud system operates based on both UDP and TCP protocols. For our first case study, due to confidentiality concerns, we have built our own cloud system and made it publicly available instead of using our partner's system directly. The second case study involves an industrial,  cloud-native IoT connected vehicle system that monitors sensors located on connected vehicles and operates based on  the MQTT protocol. In this section, we refer to the first case study system as the \emph{the UDP/TCP cloud system} and the second one
as \emph{the MQTT cloud system}.
To enhance readability and conciseness in the remainder of this section, we will use the terms ``our simulator'' or ``the simulator'' to denote a simulator that has been derived from an \lang\ specification.
The research questions (RQs) that we investigate are as follows:

\vspace*{.5em}\noindent\textbf{RQ1. (Simulator Configuration)} \emph{How can we configure our simulator so that it can, with limited computational resources, simulate a large number of IoT and edge devices?} With RQ1, we examine which simulator platform option(s) -- among Native, Docker and Virtual Machine as envisaged by the conceptual model of Figure~\ref{fig:metamodel} -- would allow one to maximize the number of instantiated IoT and edge devices, thereby best supporting our motivating use case  in Section~\ref{sec:usecase}. In particular, we require the whole simulator to run on  a single machine (laptop). This requirement is in line with the needs of our partner; they need the simulator machine to be portable so that it can be brought to different sites and connected to different networks.  In addition, we examine the impact of the speed value for edge devices on the performance of the simulator. The simulator performance is assessed by measuring (a) packet loss at the machine that hosts the simulator and  (b) packet transmission times. Since the focus of RQ1 is on the simulator, our analysis is agnostic to the choice of the cloud system. For RQ1, we elect to use the UDP/TCP cloud system. This cloud has lower execution time compared to the MQTT cloud, in turn allowing us to run more extensive experimentation.

\noindent\textbf{RQ2. (Scalability)} \emph{What is the impact of offset values on simulator scalability?} 
In many real-world scenarios, it is reasonable to assume that requests from edge devices are distributed more evenly, unlike the worst-case scenario explored in RQ1 where all edge devices initiate communication with the cloud simultaneously. To systematically examine our simulator's scalability under the assumption of more evenly distributed requests, RQ2 considers various ranges of offset values and assesses how widening the offset ranges affects scalability. Indirectly, RQ2 further enables us to empirically demonstrate the benefit of introducing offsets (in situations where engineers can control or have prior knowledge about offset ranges) as a way to increase the cloud's capacity to handle larger numbers of IoT and edge devices.
To answer RQ2, we use the best-performing platform identified in RQ1 and the UDP/TCP cloud.

\noindent\textbf{RQ3. (Comparison with Baselines)} 
\emph{How effective is our simulator at stress testing IoT cloud systems compared to state-of-the-art stress testing tools?}  We benchmark our simulator against two widely used, open-source baseline tools for stress testing: Apache JMeter~\cite{jmeter} and Locust~\cite{locust}. Both tools are commonly used for stress testing ~\cite{Araujo:19,Ismail:18, Lazidis:22, Hou:17, Koziolek:20,Pajooh:21}. We recall our conclusion from Section~\ref{sec:usecase}, where we determined that existing IoT edge-to-cloud simulators are inadequate for large-scale stress testing. In light of this conclusion, we choose to compare with JMeter and Locust (rather than IoT simulators) as our baselines, noting that JMeter and Locust are general-purpose, industry-strength performance testing tools. In Section~\ref{subsec:baselines}, we assess the ability of JMeter and Locust to tackle the three stress testing challenges, C1, C2, and C3, outlined in Section~\ref{sec:usecase}. In this RQ, we assess whether and how our simulation architecture is advantageous over JMeter and Locust in terms of the number of IoT and edge devices that can be successfully simulated. 
For a fair comparison, we use the UDP/TCP cloud: JMeter offers direct support for TCP, while Locust -- originally designed solely for HTTP -- can be easily adapted through existing libraries to work directly with UDP/TCP. Similar to RQ1 and RQ2, our focus in RQ3 is on assessing the performance of the simulator. We thus measure  packet loss incurred at the host machine and transmission delay times when running our simulator or the baselines tools.

\noindent\textbf{RQ4. (Usefulness)} \emph{Can our simulator be used effectively for stress testing real-world IoT cloud systems?}
We use the   best-performing simulator platform 
identified in \emph{RQ1} (among Native, Virtual Machine and Docker)  to answer  this research question. We  assess how well one can use our simulator to determine the service capacity of an IoT cloud system, i.e., the maximum number of IoT and edge devices that a cloud system can  effectively handle. For RQ4, we use both of our case studies, i.e. both the UDP/TCP and MQTT cloud systems. In contrast to the previous RQs which focused on studying and optimizing the behaviour of our simulator, RQ4 is concerned with in-the-field usage of the simulator, i.e., determining the limits of an IoT cloud under test. For RQ4, we measure packet loss at the destination machine hosting the cloud under test.

\noindent\textbf{RQ5. (Feedback from Practitioners)} \emph{How do practicing engineers perceive \lang?} Ultimately, the sustained use of any approach hinges on practitioners' perceptions of its usefulness. To answer RQ5, we examine the viewpoints of two engineers from our collaborating partner regarding \lang. These engineers have first-hand experience with \lang\ and can reflect on its application in actual projects. We gather opinions from the engineers through semi-structured interviews, exploring their backgrounds, experiences with \lang, and reflections.

\subsection{Case Studies}
\label{subsec:cloudBL}

In this section, we describe our case-study systems: The UDP/TCP cloud  and the MQTT cloud.

\subsubsection{The UDP/TCP Cloud}
Our UDP/TCP cloud, which is part of an industrial IoT network monitoring system, implements  a packet loopback that echoes any UDP or TCP packet  it receives from the edge layer back to the sender. 
This cloud system is used for measuring the quality of the network connection between the edge devices and the cloud.

To develop a simulator for stress testing this case-study system, we create two instances of the UDP/TCP cloud concept described in Section~\ref{subsec:metamodel}. One instance is configured to receive and transmit UDP packets, while the other handles TCP packets. We configure both instances by setting their IP attributes to match the cloud system's IP address. Further, we assign the port of the UDP cloud instance to correspond to the UDP socket's port, and the port of the other instance to match the port for the TCP socket. 
Since the cloud's host machine serves as the  destination for all inbound packets, for the \texttt{Record} method in our conceptual model, we use the default value of the \texttt{dstIP} parameter, which corresponds to the IP address of the cloud's host machine.

\subsubsection{The MQTT Cloud}
The IoT connected vehicle system monitors a large fleet of vehicles in real time by tracking and storing their location, speed, odometer and fuel-tank temperature.  Users can access the latest information of the fleet through a web application that flags the vehicles that exceed predefined thresholds for fuel-tank temperature or speed.  This system operates on a Kubernetes~\cite{kubernetes} cluster and follows a state-of-the-art big data processing approach, Lambda architecture~\cite{lambdaArchitecture}, which combines both real-time and offline processing capabilities. 

\begin{figure}[!t]
\begin{center}
\includegraphics[width=.75\linewidth]{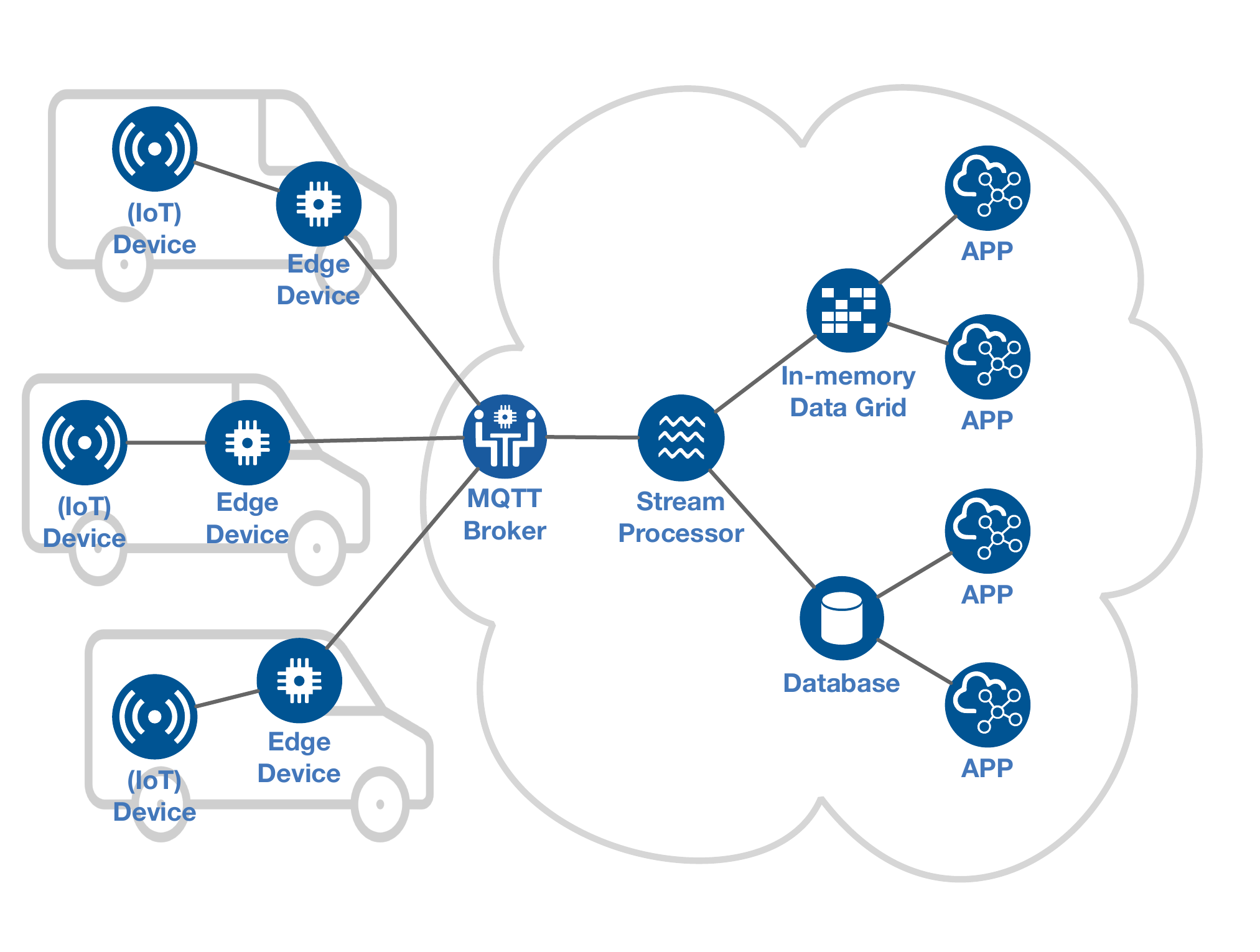}
    \vspace*{-.3cm}
    \caption{Overview of our IoT connected vehicle case study.}\label{fig:IoTArc}
    \vspace*{-.6cm}
\end{center}
\end{figure}

Figure~\ref{fig:IoTArc} provides a conceptual representation of the architecture of the IoT connected vehicle system.  This system includes an MQTT broker that accepts packets from  vehicles. 
At each time step, each vehicle sends  to the broker a packet consisting of the vehicle's GPS location, speed, odometer and fuel-tank temperature. The information in these packets enables the cloud system to  monitor and track the vehicles' position, driving direction, and speed 
as well as the temperature of the vehicles' fuel tanks.

The broker forwards these packets to a stream processor that in turn sends the packets to two different destinations: (1)~an in-memory data grid which is a fast storage system for online processing, and (2)~a persistent storage system for offline processing. Applications that need real-time data, such as  real-time diagnostics and driver-behaviour analysis, query the fast in-memory grid, while applications that do not need real-time data, such as profiling drivers' data based on machine learning, use the persistent database.

To develop a simulator for the 
IoT connected vehicle system based on our approach,  we create an edge device instance for each vehicle. Further,  we create an instance of the MQTT cloud concept, and set its \texttt{IP} attribute to the broker's IP in Figure~\ref{fig:IoTArc} and its \texttt{pubTopic} parameter to the publish topic that the broker accepts. We consider the in-memory data grid as the final destination for all packets and set the \texttth{dstIP} parameter of the \texttth{Record} method in our conceptual model to the IP address of the  in-memory data grid  in Figure~\ref{fig:IoTArc}. Note that since the IoT connected vehicle system does not send any data back to the vehicles, there is no need to set the \texttt{subTopic} parameter for our instance of the MQTT cloud concept.

\subsection{Baseline Tools for Stress Testing}
\label{subsec:baselines}

We select Apache JMeter and Locust as baseline tools to compare the performance of our simulator against. JMeter and Locust are both open-source tools for performance testing web applications. JMeter is developed by the Apache Software Foundation and written in Java. It supports various communication protocols including HTTP, FTP, TCP and SOAP/REST. Locust is built using Python and offers an easy-to-use web interface. Locust comes with built-in support for only HTTP/HTTPS.

We use JMeter and Locust to stress test our UDP/TCP cloud system. JMeter already supports TCP;
hence, we need to adapt it only for UDP. As for Locust, an adaptation to both UDP and TCP is required. To adapt JMeter, we use plugins for UDP support~\cite{jmeterplugins}; and to adapt Locust, we use Python's socket module, which provides UDP and TCP support.

We assess the performance of JMeter and Locust in tackling stress testing challenges \textbf{C1}, \textbf{C2}, and \textbf{C3} detailed in Section~\ref{sec:usecase}. Since JMeter and Locust are general-purpose tools, they do not have predefined means for capturing IoT and edge devices. Consequently, they offer no direct solution for simulating large arrays of IoT and edge devices without causing runtime bloat, i.e., challenge \textbf{C1}. Furthermore, these tools lack built-in mechanisms for addressing challenges \textbf{C2} and \textbf{C3}. To compare simulators derived from \lang\ specifications with JMeter and Locust, we adapted JMeter and Locust to symbolically represent IoT and edge devices using their parameters, as detailed in Section~\ref{subsec:settingrq3}. Through this adaptation, we effectively implemented design feature \textbf{F1} (Section~\ref{subsec:features}) into JMeter and Locust. Incorporating design features \textbf{F2} and \textbf{F3} into these tools would nonetheless necessitate a reconsideration of the tools' designs and significant re-engineering and implementation effort and was therefore not attempted. Since JMeter and Locust lack support for creating simulation nodes on virtual machines or Docker containers, when using JMeter and Locust, we run all simulation nodes natively on the host machine.

\subsection{Experiment Design}
\label{subsec:expdesign}
\begin{figure}
    \centering
    \includegraphics[width=.7\linewidth]{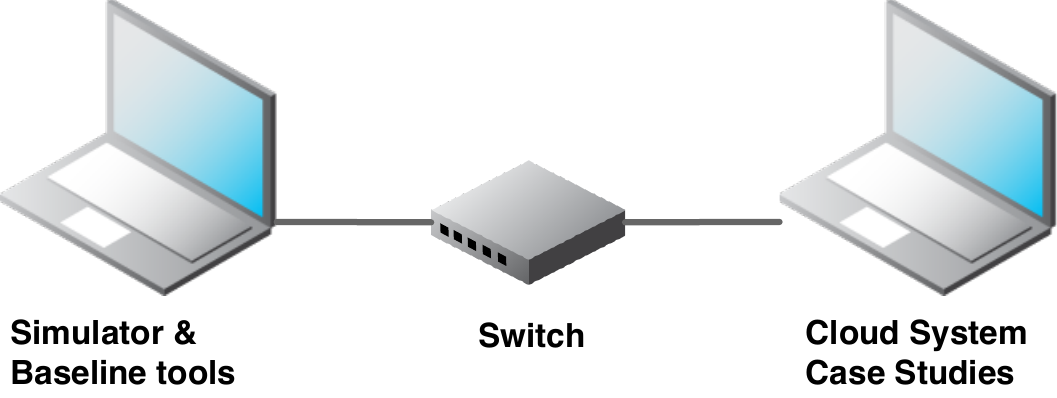}
    \caption{Physical setup for our evaluation.}
    \label{fig:evalsetup}
\end{figure}

To answer our research questions, we use two computers as shown in Figure~\ref{fig:evalsetup}:  one for running our simulator or the baseline tools, and one for running  our case studies, i.e., the UDP/TCP and MQTT cloud systems described in Section~\ref{subsec:cloudBL}.
As discussed in Section~\ref{subsec:metamodel}, in our work, we assume that the simulated edge layer and the cloud layer are connected via an ideal network with minimally low transmission time and no loss. 
We create this ideal network using a single network switch (see Figure~\ref{fig:evalsetup}). This setup, while simple, provides a high-fidelity testbed for stress testing of cloud systems, as it routes all network traffic through a switch (ideal network) rather than a real-world network that may not be as reliable or predictable.

To simulate requests from edge devices (left laptop in Figure~\ref{fig:evalsetup}), we use a  machine with a $2.5$ GHz Intel Core i9 CPU and $64$ GB of memory. For running the cloud (right laptop), we use   a machine with a $2.3$ GHz Intel Core i9 CPU and $32$ GB of memory. We connect the two machines using an unmanaged NETGEAR GS308v3 Gigabit Ethernet switch. Below, we describe the experimental setup for RQ1-RQ4 and the interview design for RQ5.

\subsubsection{Experimental Setup for RQ1}
\label{subsec:settingrq1}

\begin{table}
\caption{Parameters for the RQ1 experiments}
\label{tab:parameters-rq1}
\begin{center}
\scalebox{0.9}{
\begin{tabular}{p{4.5cm}| p{4cm}}
\toprule
duration & 2s \\
\hline
 time step   &  0.5s \\
 \hline   
\# of IoT devices per edge device  & 100 \\
\hline
\# of edge devices per sim node  & 10 \\
\hline
(IoT) Device.period&  1  \\
\hline
(IoT) Device.payload & 8B \\
\hline
Edge Device.workload &  0 \\
\hline
memory  & 2G \\
\hline
CPU & 4 \\
\hline 
Edge Device.speed & Max, 500, 250, 167, 125 \\
\hline
Platform  & Native \& VM \& CDC \& UDC\\
\bottomrule
\end{tabular}}
\end{center}
\end{table}

We conduct experiments to answer \textbf{RQ1} using the configuration parameters shown in Table~\ref{tab:parameters-rq1}. We choose a simulation duration of $2$s and a time step of $0.5$s (i.e., four steps in each simulation). Our preliminary experiments showed that the simulation duration of $2$s with four steps are sufficient to observe the behaviour of the simulator and the differences among different simulators' configurations. 
For the device payload size, we follow the recommendation of Cheetah Networks, and set it to $8$B.
We set the period of each IoT device to $1$ to maximize the number of messages sent in each time step. We assign four CPU cores and $2$G of memory to each virtual machine and constrained Docker hosting edge devices, in accordance with the computational resources typical of low-cost edge devices~\cite{Carmona:22}. We vary the number of simulation nodes from 8 to 12, corresponding to simulating 8,000 to 12,000 IoT devices. This range, in addition to  addressing RQ1, enables us to compare our simulators' performance with the results reported in Section~\ref{sec:usecase}. There, we employed two state-of-the-art edge-to-cloud simulation tools, EMU-IoT and Fogbed, to simulate 10,000 IoT devices.

For edge devices, we eliminate edge-computing time by setting \texttth{edge.workload} to zero, noting that our goal is maximizing communication with the cloud. We set the number of  IoT devices per edge device to $100$ and the number of edge devices per simulation node to $10$. That is, each simulation node captures the behaviour of $1000$ IoT devices. For the purpose of RQ1 experiments, we vary the number of simulation nodes to determine which simulator platform enables capturing more IoT devices. Bundling $1000$ IoT devices into each simulation node allows us to increase the simulator load by $1000$ as we increase the number of simulation nodes. 
 For the speed of edge devices, we insert a gap time of $0$ms, $1$ms, $2$ms, $3$ms and $4$ms between the consecutive messages sent, resulting in the edge device speed to be set, respectively, to MAX, $500$, $250$, $167$ and $125$.  Note that further decreasing the speed (i.e., increasing the gap time) leads to inconsistency in our experimental setup since each edge device needs to be able to send 100 messages in a single time step. If the gap time is $5$ms, then the total gap time for $100$ messages is $0.5$s. This is the same as the time step, so there is no time left for sending messages. 
We execute each configuration of our simulator for four platform options: Native, Virtual Machine (VM), Constrained Docker (CDC) and Unconstrained Docker (UDC).

In total, to answer \textbf{RQ1}, we perform $100$ experiments (\# of platforms = $4$) $\times$ (\# of speeds = $5$) $\times$ (\# of sim nodes = $5$) . We repeat each experiment $10$ times to account for random variation. 

\subsubsection{Experimental Setup for RQ2}
To address \textbf{RQ2}, we examine how different values for  the \texttt{offsetRange} parameter of simulation nodes would impact the scalability of our simulator. 
We use the best-performing platform from RQ1 alongside the UDP/TCP cloud for testing. Table~\ref{tab:parameters-rq2} shows the parameters specific to RQ2. For the parameters that are not listed in Table~\ref{tab:parameters-rq2}, we use the same values as those in  RQ1 (see Table~\ref{tab:parameters-rq1}). We vary the \texttt{offsetRange} attribute of simulation nodes by using the following values: $0\%$, $20\%$, $40\%$, $60\%$, and $80\%$. We set the value of the \texttt{speed} attribute to MAX. The impact of changing  
\texttt{speed} is addressed by the RQ1 results.  For the RQ2 experiments, we  specifically study the impact of changing the value of \texttt{offsetRange} on the performance of the simulator.
To illustrate the impact of \texttt{offSetRange}, we set the number of simulation nodes to range from $15$ to $19$. We increase the number of simulation nodes compared to those for RQ2 to demonstrate how offset variables enhance simulation capacity.

In total, to answer \textbf{RQ2}, we perform $50$ experiments (\# of platforms = $2$) $\times$ (\# of offsetRange = $5$) $\times$ (\# of sim nodes = $5$). We repeat each experiment $10$ times to account for random variation.

\begin{table}
\caption{Parameters for the RQ2 experiments}
\label{tab:parameters-rq2}
\begin{center}
\scalebox{0.9}{
\begin{tabular}{p{4.5cm}| p{4.5cm}}
\toprule
Edge Device.speed  & MAX\\
\hline
sim node(SN).offsetRange & 0, 20\%, 40\%, 60\%, 80\% \\
\bottomrule
\end{tabular}}
\end{center}
\end{table}

\subsubsection{Experimental Setup for RQ3}
\label{subsec:settingrq3}
To answer \textbf{RQ3}, we employ JMeter and Locust as baseline tools for conducting stress testing with the UDP/TCP cloud. For our simulator, we use the same setup as that used for RQ1 (see Table~\ref{tab:parameters-rq1}). 
In line with RQ1, we execute each baseline tool for a duration of $2$ seconds. To achieve this,  we set the  \texttt{duration} parameter of JMeter and the \texttt{run-time} parameter  of Locust to $2$. Neither of the tools explicitly support the concept of time step. We implement a custom variable in the scripts running the tools to mimic the time step and set that variable to $0.5$s. For both tools, we represent each edge device as a single \texttt{thread}.  While we can organize the threads representing edge devices into groups, such thread groups cannot act as simulation nodes because they cannot run independently on virtual machines or Dockers. Thus, for both baselines, we can only run all the edge devices  (threads) as native on the host machine. For JMeter, we employ the \texttt{loop count} parameter as a representation of the number of IoT devices associated with an edge device. We set  \texttt{loop count} to $100$. For Locust, we initiate an instance of a created class inherited from the original \texttt{user} class provided by Locust to represent an IoT device, and configure the \texttt{user count} (i.e., the number of instantiations) to be $100$.
For both tools, we use the same payload as that used for \hbox{RQ1, i.e., $8$B.}
We initially use the same range for simulation nodes as in RQ1, i.e., $8$ to $12$. We then modify the number of simulation nodes to 4 for JMeter and Locust, while increasing it to 15 for our simulator. This aims to determine the maximum number of nodes that JMeter, Locust, and our simulator can handle without packet drops on the simulator side.

In total, to answer \textbf{RQ3}, we perform $18$ experiments (\# of baselines = $2$) $\times$ (\# of sim nodes = $9$) with the baseline tools, perform $2$ experiments (\# of platforms = $2$) $\times$ (\# of sim nodes = $1$) and reuse $2$ experiments  (\# of platforms = $2$) $\times$ (\# of sim nodes = $1$) with our simulator. We repeat each experiment $10$ times to account for random variation.

\subsubsection{Experimental Setup for RQ4}
To address \textbf{RQ4},  we
use our simulator to determine the service capacity of both the UDP/TCP cloud and  the MQTT cloud systems. For the UDP/TCP cloud, we consider the experimental setup used for RQ1 (see Table~\ref{tab:parameters-rq1}), except that we consider only the platform option that, based on the results of $\textbf{RQ1}$, is capable of successfully executing on a single machine and sending all messages to the UDP/TCP cloud.

\begin{table}
\caption{Parameters for the RQ4 experiments with the MQTT cloud.}
\label{tab:parameters-rq4-2}
\begin{center}
\scalebox{0.9}{
\begin{tabular}{p{4.5cm} | p{4cm}}
\toprule
duration &  10m \\
\hline
time step   &  0.1s  \\
 \hline
 (IoT) Device.period &   1\\
\hline
(IoT)Device.payload &  166B  \\
\hline
Edge Device.workload &   0  \\
\hline
sim node (SN).offsetRange & 0 \\
\hline
memory & 4G\\
\hline
CPU & 4  \\
\bottomrule
\end{tabular}}
\end{center}
\end{table}

Table~\ref{tab:parameters-rq4-2} shows the parameters used in our experiments with the MQTT cloud for \textbf{RQ4}. Recall from Section~\ref{subsec:cloudBL} that the MQTT cloud case study represents a connected vehicle system where one edge device is installed on each vehicle.  
We set the simulation duration to $10$ minutes, which is close to the average transport time for vehicles (e.g., trucks) to travel per trip (from one location to another location by motor vehicles)~\cite{Hoh:07}.  We set the time step to $0.1$s for our simulation, which corresponds to a data transmission rate of $10$Hz. 
This decision aligns with existing research in connected vehicles~\cite{Sonklin:19}, which suggests that data transmission rate typically ranges from $1$ to $10$Hz, reflecting real-world scenarios. That is, vehicles  send one to ten messages per second to roadside infrastructures or cloud systems. Specifically, a frequency of $10$Hz is often used for safety messages and location data exchanges within cooperative \hbox{awareness services~\cite{Sonklin:19}.}

Each packet in this case study contains all the vehicle's information, i.e., Vehicle ID, GPS location, speed, odometer, and fuel-tank temperature.  The payload of each IoT device is configured as a JSON string. The payload size is on average $166$B and varies slightly (standard deviation of $\approx$$4$ bytes) depending on the specific values included in the JSON string. In our setup, we set the period of IoT devices to $1$, indicating that they generate data at a frequency of one per simulation step; this is to maximize the number of messages sent in each time step. 
We set the \texttt{edge.workload} parameter to $0$ because edge devices in this system do not perform edge computing tasks.
To simulate the behaviour of edge devices running on vehicles, we allocate $4$ CPU cores and $4$G of memory to the platform with constraint computational resources. These settings align with the current embedded hardware commonly used in vehicles. For the rest of the parameters not included in Table~\ref{tab:parameters-rq4-2}, we use the same values as those in RQ1.

To determine the capacity of each of our case-study cloud systems, we vary the number of simulation nodes 
from $29$ to $34$ and find the threshold at which packet loss is observed on the cloud side.

For \textbf{RQ4}, we reuse $10$ experiments  (\# of platforms = $2$) $\times$ (\# of sim nodes = $5$) from RQ1 with the UDP/TCP cloud and perform $12$ experiments (\# of sim nodes = $6$) $\times$ (\# of platforms = $2$) with the MQTT cloud.  Each experiment is repeated $10$ times to account for randomness. 

\subsubsection{Interview Design for RQ5}

To address \textbf{RQ5}, we conduct semi-structured interviews with two engineers from our industry partner. The interview questions are divided into five parts. The first part explores the background of the interviewees, focusing on their work experience in the IoT domain. The second part asks about the interviewees' experience in quality assurance, specifically the validation, verification, and testing of IoT cloud systems. The third part queries interviewees about alternative tools employed for similar purposes as \lang\ and the relative pros and cons of \lang\ compared to past practices. The fourth part gathers insights into the costs related to learning and using \lang. Finally, the fifth part elicits interviewees' perspectives on the usefulness of \lang\ and potential future enhancements.  Both the interview questions and transcripts are provided in our supplementary material.

\subsection{Metrics}
\label{subsec:metric}
We measure two main metrics in our experiments: (1)~the number of packets that are dropped during the simulation, and (2)~the packet transmission time, i.e., the amount of time it takes for the packets to be transmitted from the simulator to the cloud and be processed by the cloud. In the setup of Figure~\ref{fig:evalsetup}, packets can be dropped (1) on the simulator side when the simulator fails to send all the messages it is expected to send, (2) on the switch (network) side when the switch fails to handle the traffic that needs to pass through it, e.g., due to congestion, or (3) on the cloud side, when the cloud fails to receive all the packets that have reached its host computer, or fails to send responses for the packets it has received. We separately measure packet drop at the three above points  and refer to them respectively as \emph{SimDrop}, \emph{NetDrop} and \emph{CloudDrop}. Note that in order to measure these values, we have to keep track of  packet counts at the level of the simulator, the cloud and also the network adapters on both host machines. To count the number of packets sent from and received by the network adapters of the host machines of the simulator and the cloud, we use the Wireshark tool~\cite{Wireshark} which is the world's most widely used network protocol analyzer. Using Wireshark, we have confirmed that, in our experiments,  the packet drop at the network adaptors is zero or negligible (less than five packets out of hundreds of thousands). In addition,  due to our utilization of a switch with a data transfer rate of up to $1$ Gbps per port, capable of managing the transmission of more than $8,000$ simulation nodes per step without surpassing its capacity, to connect the computers, the packet drop for the network is zero as well. Hence, we  do not report \emph{NetDrop} as it is always zero for our experiments.
The packet drop values measured at the simulator (i.e., \emph{SimDrop}) and at the cloud (i.e, \emph{CloudDrop}) as well as \emph{packet transmission time} (as defined above) are thus the only measures that are of interest for capturing the behaviour of the simulator and the cloud.

Among these measures, \emph{SimDrop} determines if the simulator is able to scale and that it does not fail under its own load. If there is packet drop by the simulator, it means that the simulator is overwhelmed. The \emph{CloudDrop} and the packet transmission time (\emph{TransTime}, for short) determine how well the cloud is able to handle the messages it receives from the edge.

\subsection{Results}

\begin{table}[hbtp]
 \caption{Average packet drop and  transmission time values obtained by ten runs of the experiments of RQ1. The values highlighted grey in the table are the results used to answer RQ4 for the UDP/TCP cloud system. }
   \label{tbl:rq1}
\begin{center}
\scalebox{0.8}{
\begin{tabular}{p{0.5cm}  p{1.8cm}  p{0.8cm}  p{0.8cm}  p{0.8cm}  p{0.8cm}  p{0.8cm}}
\toprule
\#SN&Speed&MAX&500&250&167&125\\
\midrule
 \multirow{16}{*}{12}&&&\textbf{CDC}&&&\\
&SimDrop&\cellcolor{lightgray}0&0&0&0.2&0\\
&CloudDrop&\cellcolor{lightgray}2314.5&0&0&0&0\\
&TransTime(ms)&\cellcolor{lightgray}6.61&0.49&0.74&0.52&0.29\\  
 &&&\textbf{UDC}&&&\\
 &SimDrop&\cellcolor{lightgray}0&0&0&0&0\\
&CloudDrop&\cellcolor{lightgray}2805.5&0&0&0&0\\
&TransTime(ms)&\cellcolor{lightgray}7.05&0.48&0.61&0.46&0.31\\  
 &&&\textbf{VM}&&&\\
 &SimDrop&42.6&3254.4&2922.2&4383.9&6294.2\\
&CloudDrop&28.1&0&0&0&0\\
&TransTime(ms)&0.47&0.44&0.27&0.28&0.28\\  
 &&&\textbf{Native}&&&\\
 &SimDrop&25051.1&25563.8&25958.1&26089.8&26312.4\\
&CloudDrop&0&0&0&0&0\\
&TransTime(ms)&0.35&0.26&0.26&0.26&0.26\\  

     \midrule
 \multirow{16}{*}{11}&&&\textbf{CDC}&&&\\
 &SimDrop&\cellcolor{lightgray}0&0&0&0&0\\
&CloudDrop&\cellcolor{lightgray}3246.1&70.5&0&0&0\\
&TransTime(ms)&\cellcolor{lightgray}10.57&4.31&1.32&0.33&0.28\\  
&&&\textbf{UDC}&&&\\
 &SimDrop&\cellcolor{lightgray}0&0&0&0&0\\
&CloudDrop&\cellcolor{lightgray}3741.2&0&0&0&0\\
&TransTime(ms)&\cellcolor{lightgray}8.11&1.30&0.87&0.37&0.30\\  
 &&&\textbf{VM}&&&\\
 &SimDrop&71&119.10&1272.9&2407.3&4410.3\\
&CloudDrop&0&0&0&0&0\\
&TransTime(ms)&0.38&0.28&0.28&0.28&0.28\\  
 &&&\textbf{Native}&&&\\
 &SimDrop&21022.4&21543.4&21980.3&22060.8&22599.6\\
&CloudDrop&0&0&0&0&0\\
&TransTime(ms)&0.26&0.26&0.28&0.33&0.30\\  

    \midrule
 \multirow{16}{*}{10}&&&\textbf{CDC}&&&\\
 &SimDrop&\cellcolor{lightgray}0&0&0&0&0\\
&CloudDrop&\cellcolor{lightgray}1517.2&0&0&0&0\\
&TransTime(ms)&\cellcolor{lightgray}9.01&1.74&0.70&0.29&0.27\\  
 &&&\textbf{UDC}&&&\\
 &SimDrop&\cellcolor{lightgray}0&0&0&0&0\\
&CloudDrop&\cellcolor{lightgray}296.6&0&0&0&0\\
&TransTime(ms)&\cellcolor{lightgray}3.52&1.00&0.61&0.34&0.29\\  
 &&&\textbf{VM}&&&\\
 &SimDrop&69.1&423.7&178.3&710.1&2322.9\\
&CloudDrop&0&0&0&0&0\\
&TransTime(ms)&0.31&0.28&0.28&0.50&0.28\\  
 &&&\textbf{Native}&&&\\
 &SimDrop&17252.6&17546.1&18103&18128.8&18537\\
&CloudDrop&0&0&0&0&0\\
&TransTime(ms)&0.26&0.26&0.29&0.32&0.26\\

    \midrule
 \multirow{16}{*}{9}&&&\textbf{CDC}&&&\\
 &SimDrop&\cellcolor{lightgray}0&0&0&0&0\\
&CloudDrop&\cellcolor{lightgray}768&0&0&0&4.3\\
&TransTime(ms)&\cellcolor{lightgray}4.71&1.38&0.62&0.28&0.62\\  
 &&&\textbf{UDC}&&&\\
 &SimDrop&\cellcolor{lightgray}0&0&0&0&2.2\\
&CloudDrop&\cellcolor{lightgray}0&0&0&0&0\\
&TransTime(ms)&\cellcolor{lightgray}0.62&0.61&0.51&0.30&0.28\\  
 &&&\textbf{VM}&&&\\
 &SimDrop&0&107.7&79.6&50.4&844.8\\
&CloudDrop&0&0&0&0&0\\
&TransTime(ms)&0.29&0.26&0.26&0.30&0.28\\  
 &&&\textbf{Native}&&&\\
 &SimDrop&13239.5&13834&13951.9&14209.4&14136.1\\
&CloudDrop&0&0&0&0&0\\
&TransTime(ms)&0.26&0.28&0.28&0.26&0.26\\

    \midrule
 \multirow{16}{*}{8}&&&\textbf{CDC}&&&\\
 &SimDrop&\cellcolor{lightgray}0&0&0&0&0\\
&CloudDrop&\cellcolor{lightgray}0&0&0&0&0\\
&TransTime(ms)&\cellcolor{lightgray}0.99&0.98&0.35&0.28&0.26\\  
 &&&\textbf{UDC}&&&\\
 &SimDrop&\cellcolor{lightgray}0&0&0&0&0\\
&CloudDrop&\cellcolor{lightgray}0&0&0&0&0\\
&TransTime(ms)&\cellcolor{lightgray}0.41&0.49&0.41&0.29 & 0.27\\  
 &&&\textbf{VM}&&&\\
 &SimDrop&26.8&0.9&0&0&100.7\\
&CloudDrop&0&0&0&0&0\\
&TransTime(ms)&0.27&0.27&0.27&0.26&0.30\\  
 &&&\textbf{Native}&&&\\
&SimDrop&9323.7&9858.3&9935.7&10176.4&10168\\
&CloudDrop&0&0&0&0&0\\
&TransTime(ms)&0.26&0.26&0.30&0.26&0.29\\  

\bottomrule
\multicolumn{7}{l}{ * VM: Virtual Machine}\\
\multicolumn{7}{l}{* CDC: Constrained Docker Container}\\
\multicolumn{7}{l}{* UDC: Unconstrained Docker Container}
\end{tabular}}
\end{center}
\end{table}

\subsubsection{RQ1.(Simulator Configuration)}
Table~\ref{tbl:rq1} shows the average values for SimDrop, CloudDrop, and packet transmission time obtained by running our simulator using the parameters in Table~\ref{tab:parameters-rq1}. 
We perform RQ1 experiments for $8$ to $12$ simulation nodes that correspond to simulating  $8,000$ to  $12,000$ IoT devices.
By considering a range of simulation nodes, we validate the consistency of our results, ensuring that the results remain robust for different numbers of simulated IoT devices.  As discussed in Section~\ref{subsec:settingrq1}, for RQ1, we run our simulator using four different platform options, i.e., Native, VM,  CDC and UDC. In the Native option, we run all the edge devices in parallel on the simulator host machine without grouping them into separate simulation nodes. In contrast, when we use VM, CDC, UDC, we group every ten edge devices into a simulation node and run each simulation node on a separate VM, CDC or UDC. For each VM and each CDC, we use the CPU and memory sizes specified in Table~\ref{tab:parameters-rq1}.

As shown in Table~\ref{tbl:rq1}, there is substantial packet drop on the simulator side when we use the Native option. This indicates that in this architecture where edge devices are not grouped into simulation nodes, the simulator is simply unable to send all the packets it is supposed to send. 
This highlights the importance of the hierarchical structure provided by the \texttth{Simulation Node} concept in \lang. Similarly, the VM option, while generating SimDrop values lower than those of the Native option, is unsuitable for stress testing as it still leads to non-negligible packet drop on the side of the simulator.

For the simulation configurations executed on CDC and UDC, SimDrop values are zero (or negligible). For these configurations, the cloud side drops packets, particularly  when edge devices send their packets  at once and without having any gap time in between (i.e., when the edge device speed is MAX). By inserting a gap time on the side of edge devices (i.e., lowering the speed), the cloud will receive the packets at a lower rate and can avoid  packet drop.

Note that transmission-time values for the Native and VM should be discarded. This is because, for these options, there is a high packet drop at the simulator, and hence, considerably fewer packets are transmitted from the simulator to the cloud and vice versa. Therefore,  the transmission times are computed for a smaller portion of packets, leading to unrealistically small values. For the CDC and UDC options where the simulator is able to complete its task without packet drop, the transmission time decreases as we reduce the speed of the edge devices. This trend is expected, since by reducing the speed, we avoid the cloud side from being overwhelmed, hence reducing the packet transmission time.  

\vspace*{.3cm}
\resq{The answer to \textbf{RQ1} is that  our simulator can be tuned such that it can, on a conventional laptop, simulate the sending of $24,000$ packets per second without packet drop. Further, our results show the importance of grouping edge devices into simulation nodes to manage resource usage on the simulator's host machine and thus enable the simulator to scale as much as possible. In our experimental setup,  containerization (through Dockers) turned out to be the best option for scaling. Among the main benefits of \lang\ are its support for simulation nodes and its ability to abstract from platform details so that engineers can explore different platform options.}
\vspace*{.3cm}

\begin{table}[t]
\begin{center}
\caption{Average packet drop and  transmission time values obtained by ten runs of the experiments of RQ2.}
\label{tbl:rq2}

\scalebox{0.8}{
\begin{tabular}{
p{0.5cm} p{1.8cm} p{0.8cm} p{0.8cm} p{0.8cm} p{0.8cm} p{0.8cm}}
\toprule
\#SN&OffsetRange&80\%&60\%&40\%&20\%&0\\
\midrule
\multirow{8}{*}{19}&&&\textbf{CDC}&&&\\
& SimDrop &  0 &  0 &  0 &  0 &  307.8 \\
& CloudDrop &  2971.5 &  3161.2 &  3363.3 &  3503.0 &  3697.1 \\
& TransTime(ms) &  8.13 &  8.39 &  8.65 &  8.74 &  10.04 \\
&&&\textbf{UDC}&&\\
& SimDrop &   0 &  0 &  0 &  0 &  369.4  \\
& CloudDrop &  2702.3 &3076.7 &3178.0 &3590.3 &3486.4  \\
& TransTime(ms) &   8.49 &  8.75 &  9.02 &  9.17 &  10.54  \\

\midrule
\multirow{8}{*}{18}&&&\textbf{CDC}&&\\
& SimDrop &  0 &  0 &  0 &  0 &  263.5 \\
& CloudDrop &   2642.1 &  2810.7 &  2990.2 &  3114.8 &  3308.9  \\
& TransTime(ms) &   7.89 &  8.21 &  8.40 &  8.29 &  9.48 \\
&&&\textbf{UDC}&&&\\
& SimDrop &  0 &  0 &  0 &  0 &  316.3   \\
& CloudDrop &  3136.0 &2904.2 &3480.8 &3236.9 &3576.1  \\
& TransTime(ms) &   8.41 &  8.16 &  8.68 &  8.70 &  9.94  \\

\midrule
\multirow{8}{*}{17}&&&\textbf{CDC}&&\\
& SimDrop &  0 &  0 &  0 &  0 &  206.3  \\
& CloudDrop &   2793.4 &  2971.7 &  3161.4 &  3293.5 &  3471.1 \\
& TransTime(ms) &  7.43 &  7.31 &  7.77 &  7.84 &  8.91 \\
&&&\textbf{UDC}&&\\
& SimDrop &   0 &  0 &  0 &  0 &  247.6 \\
& CloudDrop &  3024.0 &2996.2 &2908.5 &3261.2 &3362.6  \\
& TransTime(ms) &  7.81 &  8.06 &  8.31 &  8.23 &  9.35  \\

\midrule
\multirow{8}{*}{16}&&&\textbf{CDC}&&\\
& SimDrop &   0 &  0 &  0 &  0 &  161.3 \\
& CloudDrop &   2492.2 &  2651.3 &  2820.6 &  2938.2 &  3115.8  \\
& TransTime(ms) &  6.98 &  7.11 &  7.31 &  7.39 &  8.35  \\
&&&\textbf{UDC}&&\\
& SimDrop &  0 &  0 &  0 &  0 &  193.6   \\
& CloudDrop &  2923.8 &2704.0 &3125.9 &3328.3 &3458.2  \\
& TransTime(ms) &  7.21 &  7.43 &  7.67 &  7.75 &  8.76  \\

\midrule
\multirow{8}{*}{15}&&&\textbf{CDC}&&\\
& SimDrop &    0 &  0 &  0 &  0 &  153.4 \\
& CloudDrop &    2336.5 &  2485.7 &  2644.4 &  2754.6 &  2916.4 \\
& TransTime(ms) &   6.66 &  6.79 &  7.02 &  6.93 &  7.81 \\
&&&\textbf{UDC}&&\\
& SimDrop &   0 &  0 &  0 &  0 &  184.1 \\
& CloudDrop &  2182.5 &2465.7 &2528.0 &2644.7 &3524.9 \\
& TransTime(ms) &    6.97 &  7.09 &  7.31 &  7.27 &  8.20  \\
\bottomrule
\end{tabular}}
\end{center}
\end{table}

\subsubsection{RQ2. (Scalability)}
We address RQ2 by analyzing how the \texttt{offsetRange}  parameter affects our simulator's scalability. We note that higher values of \texttt{offsetRange} mean that the edge devices' offsets can vary within a larger range. Thus, the edge devices' start times are more spread out within a time step.  The experiments for RQ2 are performed for $15$ to $19$ simulation nodes that correspond to simulating  $15,000$ to  $19,000$ IoT devices. The results, presented in Table~\ref{tbl:rq2}, show that by increasing the value of \texttt{offsetRange}, the simulator is able to send more messages successfully. Specifically, as shown in the table,  SimDrop becomes zero when the \texttt{offsetRange}  exceeds $20$\%. In other words, when we force all edge devices to start at the beginning of the time step, i.e., all offset values are zero, our simulator incurs some drop for all the simulation nodes shown in Table~\ref{tbl:rq2}. However, when the offset values can vary within a range, the simulator successfully simulates up to $19,000$ IoT devices within half a second. This is approximately 1.3 times higher than the number of IoT devices our simulator could handle in RQ1 when the start time of edge devices is set to zero.

\vspace*{.3cm}
 \resq{The answer to RQ2 is that by introducing randomness in the start time of edge devices, our simulator can scale to at least $1.3$ times more IoT devices compared to situations where  randomness is excluded.}
\vspace*{.3cm}

\begin{table}[t]
  \caption{Average packet drop and  transmission time values obtained by ten runs of the experiments of RQ3 (Baselines).}
  \label{tbl:rq3}
  \begin{center}
      \scalebox{0.8}{
      \begin{tabular}{p{0.5cm} p{1.8cm}p{0.8cm} p{0.8cm}}
\toprule
\#SN&Baselines&JMeter&Locust\\
\midrule
 \multirow{3}{*}{12}&SimDrop&11461.7&9985.5\\
&CloudDrop&1.8&1.8\\
&TransTime(ms)&0.67&0.42\\

     \midrule
 \multirow{3}{*}{11}&SimDrop&4594.6&3984.4\\
&CloudDrop&4.4&4.4\\
&TransTime(ms)&0.46&0.30\\

    \midrule
 \multirow{3}{*}{10}&SimDrop&2789.6&2421.2\\
&CloudDrop&0&0\\
&TransTime(ms)&0.40&0.36\\

    \midrule
 \multirow{3}{*}{9}&SimDrop&687.6&594.9\\
&CloudDrop&0&0\\
&TransTime(ms)&0.33&0.36\\

    \midrule
 \multirow{3}{*}{8}&SimDrop&559.4&484.1\\
&CloudDrop&0&0\\
&TransTime(ms)&0.35&0.33\\
\midrule
...\\
\midrule
 \multirow{3}{*}{5}&SimDrop&213.2&189.7\\
&CloudDrop&0&0\\
&TransTime(ms)&0.31&0.29\\
\midrule
 \multirow{3}{*}{4}&SimDrop&0&0\\
&CloudDrop&0&0\\
&TransTime(ms)&0.28&0.26\\
\bottomrule
\end{tabular}}
\end{center}
\end{table}

\begin{table}[t]
  \caption{Average packet drop and  transmission time values obtained by ten runs of the experiments of RQ3 (Our Simulator).}
  \label{tbl:rq3-2}
  \begin{center}
      \scalebox{0.8}{
      \begin{tabular}{p{1cm} p{2cm} p{1cm} p{1cm} }

\toprule
\#SN&Our Simulator& CDC&UDC\\
\midrule
    
 \multirow{3}{*}{15}&SimDrop& 153.4 & 184.1\\
&CloudDrop&2916.4&3524.9\\
&TransTime(ms)&7.81&8.20\\

     \midrule
 \multirow{3}{*}{14}&SimDrop&0&0\\
&CloudDrop&2834.1&3116.4\\
&TransTime(ms)&6.53&7.41\\
\bottomrule
\end{tabular}}
  \end{center}
\end{table}

\subsubsection{RQ3. (Comparison with Baselines)}
To answer \textbf{RQ3}, we apply our baseline tools Apache JMeter and Locust  to the UDP/TCP cloud using the settings discussed in Section~\ref{subsec:settingrq3}. The average SimDrop, CloudDrop and packet-transmission time results obtained from the baseline tools are shown in Table~\ref{tbl:rq3}. We compare the results obtained from the baselines, shown in Table~\ref{tbl:rq3}, with the results from our simulator, shown in Table~\ref{tbl:rq1}. The results were obtained by running the baselines and our simulator under identical settings as detailed in Section~\ref{subsec:settingrq3}.   As Table~\ref{tbl:rq3} shows, there is significant packet drop on the simulator side in all the experiments conducted with JMeter and Locust. Compared to the results in Table~\ref{tbl:rq1}, both JMeter and Locust exhibit lower SimDrop values than those of the Native platform. Although the baselines perform better than the Native platform, both perform significantly worse than the CDC, UCD, and VM platforms.

To determine the  number of simulation nodes our simulator can handle without packet drops on the simulator side, we conduct extended experiments where we gradually increase the number of simulation nodes until we observe non-zero SimDrop. Table~\ref{tbl:rq3-2} shows that SimDrop occurs in our simulator when we have $15$ or more simulation nodes.  In other words, our simulator can successfully simulate up to $14$ simulation nodes in one time step (i.e., $0.5$s), which is equivalent to $28,000$ IoT devices per second. For JMeter and Locust, we decrease the number of simulation nodes until the value of SimDrop becomes zero or negligible. The part below the row of ``...'' in Table~\ref{tbl:rq3} shows that both JMeter and Locust lead to non-negligible SimDrop with $5$ or more simulation nodes. This means that JMeter and Locust can successfully simulate up to $4$ simulation nodes in a single time step, corresponding to $8,000$ IoT devices per second. Therefore, our simulator can simulate $3.5$ times more IoT devices than JMeter and Locust.

We note that the transmission-time and CloudDrop values are not used for comparison. These values are relatively low or negligible for the baselines. This is because, due to the high SimDrop, the baselines can transmit only a small portion of their packets. This leads to artificially lower transmission time and CloudDrop values when compared to the CDC and UDC platforms, which can transmit all their packets. Hence, these values cannot be used as a reliable basis for comparison.

\vspace*{.2cm}
\resq{The answer to \textbf{RQ3} is that our simulator, when utilizing the CDC and UDC platforms, outperforms the baseline tools, namely JMeter and Locust, by successfully simulating  $3.5$ times more IoT and edge devices compared to what can be achieved using the baseline tools.}
\vspace*{.5cm}

\subsubsection{RQ4. (Usefulness)} 
We answer RQ4 using both the UDP/TCP cloud and the MQTT cloud systems. 
Regarding the UDP/TCP cloud, we are interested in the the maximum number of IoT devices that our UDP/TCP cloud can serve without significant CloudDrop. To find this number, we recall the RQ1 results highlighted grey in Table~\ref{tbl:rq1}.  These results show that the UDP/TCP cloud reaches its limit, i.e., starts dropping packets,  when the number of simulation nodes is $9$ or higher for the CDC platform and $10$ or higher  for the UDC platform.  Therefore, the threshold  for the service capacity of our UDP/TCP cloud is $8$ simulation nodes, since it is the highest number of simulations nodes that does not lead to significant CloudDrop for both platforms. This means that our UDP/TCP cloud can handle -- without packets being dropped by the cloud -- $8,000$ IoT devices sending messages every $0.5$s or $16,000$ IoT devices sending packets every second.

\begin{table*}[t]
\begin{center}
\caption{Average packet drop and packet transmission time values obtained by ten runs of the experiments of RQ4 (stress testing MQTT system).}
  \label{tbl:rq4-2}
\scalebox{0.9}{\begin{tabular}{p{1.8cm}
p{1.6cm}p{1.6cm} p{1.6cm}p{1.6cm} p{1.6cm} p{1.6cm}}\\
\toprule  
    Metrics & \#Vehicles=340 & \#Vehicles=330 & \#Vehicles=320 & \#Vehicles=310 & \#Vehicles=300 &
    \#Vehicles=290\\
    \midrule
    &&& \textbf{CDC} &&& \\
    CloudDrop & 17050.8 & 16431.4 & 0 & 0 & 1253.1 & 0\\
    TransTime(ms) & 6.37 & 6.49 & 5.94 & 5.78 & 5.63 & 5.66 \\
    &&& \textbf{UDC} &&&\\
    CloudDrop & 16538.1 & 16153.5 & 0 & 0 & 970.2 & 0 \\
    TransTime(ms) & 6.51 & 6.33 & 5.87 & 5.61 & 5.27 & 5.11\\
    \bottomrule
\end{tabular}}
\end{center}
\end{table*}

Table~\ref{tbl:rq4-2} presents the  CloudDrop and transmission-time values obtained by applying our simulator to the MQTT cloud. Since the SimDrop values are consistently zero, these values are not included in the table. The results shown in Table~\ref{tbl:rq4-2} are obtained when the number of simulated  vehicles is set to $290$, $300$, $310$, $320$, $330$ and $340$, respectively. 
Recall from Section~\ref{subsec:expdesign} that for the MQTT cloud, each vehicle acts as one edge device with one associated IoT device, and further, each vehicle sends one packet with a payload of $166$B every $100$ms. As shown in Table~\ref{tbl:rq4-2}, CloudDrop  is significant when the number of vehicles  is more than $320$. Note that even for $300$ vehicles, CloudDrop is non-zero, i.e., $1,253.1$ and $970.2$ when we used our simulator with the CDC and UDC platforms, respectively. But this level of CloudDrop is $0.06\%$ (negligible) considering that we send a total number of $1,800,000$ packets in the experiments with the MQTT cloud. Hence, CloudDrop is considered $0$ or negligible when the number of vehicles is less than or equal to $320$. The results in Table~\ref{tbl:rq4-2} show that the MQTT cloud can successfully handle messages coming from up to $320$ vehicles at every $100ms$ or messages coming from up to $3,200$ vehicles per second. We note that the packet payload for the MQTT case study is 20 times larger than the packet payload for our UDP/TCP case study. Due to the packet size differences and other differences between these two systems,  the number of IoT devices that they can handle are different.

\vspace*{.3cm}
 \resq{The answer to \textbf{RQ4} is that our simulator successfully determines the service capacity of our case studies -- the UDP/TCP cloud and the MQTT cloud. Specifically, the UDP/TCP cloud has the ability to accommodate a load of $16,000$ IoT devices each sending one message every second, while the MQTT cloud can serve $3,200$ vehicles each sending one message every second.}

\subsubsection{RQ5. (Feedback from Practitioners)} 
\label{results:RQ5}
We address \textbf{RQ5} through insights gathered from the two interviewees. As mentioned earlier, the interview questions and the detailed answers of the interviewees are provided in the supplementary material. Here, we present a synthesis of the feedback received.

Both interviewees have a minimum of five years of professional experience in IoT, including the verification and testing of IoT cloud systems. Both have actively used \lang\ for over six months. Neither interviewee is a co-author.

Based on the interviews, our partner company primarily uses \lang\ for assessing QoE (Quality of Experience), conducting network stability analysis, and performing stress testing on IoT clouds. Before adopting \lang, the company relied on standard tools such as ping and iPerf~\cite{iPerf}, custom scripts, or a combination thereof. Both interviewees found \lang\ to be considerably more easily configurable for diverse testing scenarios than alternatives. One interviewee emphasized the seamless operation of simulators built using \lang, regardless of the number of simulated IoT devices.

According to the interviewees, the effort users need to invest in learning \lang\ depends on their familiarity with cloud testing and basic Unix skills. For those already acquainted, the learning curve is minimal (on the order of minutes), whereas less experienced users may require more time (on the order of a few hours), including time to acquire fundamental Unix knowledge and understand cloud testing. One interviewee rates \lang\ as very easy to use, while the other considers it generally easy with the prerequisite knowledge.

Both interviewees strongly believed that \lang\ is useful for stress testing IoT cloud systems. One noted potential savings of up to 80\% in testing effort and around 15\% in overall project effort, equating to a time-saving of three to four person-weeks in the last six months. The second interviewee cited a minimum saving of two dedicated person-weeks over the three most recent major projects.

\vspace*{.3cm}
 \resq{The answer to \textbf{RQ5} is that, based on feedback from two engineers at our partner company who have first-hand experience using \lang\ and neither of whom is a co-author, \lang\ is useful for stress testing IoT cloud systems. Compared to alternative methods used in the past, the engineers reported that \lang\ resulted in significant time savings, reducing testing effort by up to 80\%. Furthermore, they found \lang\ to be user-friendly and easily configurable for different testing scenarios.}
\vspace*{.3cm}

\subsection{Limitations and Validity Considerations}
\label{subsec:limitation&validity}
\subsubsection{Limitations}
\label{subsec:limitation}
Below, we outline the limitations of our approach, organizing them into four aspects.

\textbf{Network protocols.} \lang\ currently supports three network communication protocols: MQTT, TCP and UDP. MQTT is widely recognized as the predominant communication protocol for IoT systems~\cite{soni:17}, while TCP and UDP are among the most extensively employed transport layer protocols for networking, serving communication needs across the Internet~\cite{kumar:12}. 
We anticipate that these protocols will adequately cover the vast majority of communication scenarios between edge devices and cloud systems. If additional protocols are required in the future, updates to IoTECS and further empirical evaluation would be necessary. The impact of incorporating new protocols on the language itself would be minor, with only Rule $\epsilon 7$ in Section~\ref{subsec:dsl} being affected. Nevertheless, it would also be necessary to develop appropriate new Xtend rules for code generation and conduct further experimentation with the new protocols.

\textbf{Immutability.} Currently, simulators derived from \lang\ specifications are immutable, meaning their configuration cannot be changed at runtime. To modify the simulator, one must alter the specification, then re-generate and re-execute the simulator. Introducing support for changing a simulator's parameters such as speed, offsetRange, duration and time step would enhance the usability and applicability of our simulators.

\textbf{Analysis.} The output metrics, such as SimDrop, CloudDrop, NetDrop, and TransTime, used in our evaluation of Section~\ref{sec:eval}, are currently obtained only at the end of a simulation run. This approach to gathering metrics does not support online monitoring, especially during the typically long testing periods for IoT systems. Our work could be expanded to enable the collection of metrics in real-time during simulator execution through additional engineering effort, without affecting our overall approach.

\textbf{Simulation.} Currently, \lang\ specifications support the introduction of uniform latencies in packets sent by simulated edge devices. Industry feedback (see the supplementary material) suggests incorporating a feature that enables custom latencies to be introduced in specific generated packets to further increase the flexibility of defining communication patterns between edge devices and IoT cloud systems.

As evidenced by the feedback from our industry partner (Section~\ref{results:RQ5}), the above limitations have not impeded the adoption of \lang\ in practice, and \lang\ remains in active use by our industry partner. That being said, prospective adopters of \lang\ need to take note of the following practical implications of the current limitations:
\begin{itemize}
\item \emph{Features:} \lang\ currently only supports UDP, TCP, and MQTT. Extending support to other protocols necessitates language extensions. Similarly, enabling customization of latency patterns on a per-packet or per-group basis, if required, involves improvements to the language.
\item \emph{Usability:} The behaviour of a simulator remains fixed at runtime. As such, users must re-generate the simulator whenever they wish to adjust simulation parameters. Further, simulation results are generated only upon completion rather than being streamed progressively as they become available.
\end{itemize}

\subsubsection{Validity Considerations}
Construct and external validity are the most relevant dimensions of validity for our evaluation.

\emph{Construct Validity:} 
The main concern for construct validity is the reliable measurement of our evaluation metrics as defined in Section~\ref{subsec:metric}. These metrics are based on common practices in network and performance engineering. To accurately measure our metrics of interest, we use Wireshark~\cite{Wireshark} -- a standard and widely used network protocol analyzer.

\emph{External Validity:} We have used \lang\ to generate simulators for stress testing two industrial case studies, i.e., the UDP/TCP and MQTT cloud systems outlined in Section~\ref{subsec:cloudBL}. 
The results obtained for both case studies indicate that \lang\ can effectively meet the stress testing needs of real-world IoT cloud systems. This provides confidence about the broader applicability of \lang. That said, conducting further evaluations with additional case studies would contribute to enhancing external validity.

Furthermore, in relation to the interviews with engineers at our industry partner, it is important to note that these interviews were intended to gather feedback from firsthand use. So far, to our knowledge, only our partner company has adopted \lang. Therefore, not only is the number of interviewees small, but also our sample does not capture diverse perspectives from multiple companies. Consequently, our current interview results should not be taken to imply that our approach is necessarily useful in a broader context. Discussing broader usage requires wider-scale adoption of our results and larger interview and survey studies.

\section{Related Work}
\label{sec:related}

We compare our work with the related work in the areas of IoT simulation, IoT testing, and domain-specific languages for IoT. 
\subsection{Simulation and Testing of IoT systems}

\begin{table*}
\caption{
Comparison of \lang\ with relevant work strands in IoT.}
\label{tab:relatedwork}
\vspace*{-.5em}
\begin{center}
\renewcommand\arraystretch{1.4}
\scalebox{1.1}{
\begin{tabular}{|p{3.5cm}|p{3cm} | p{4cm}|p{1.7cm}|p{1.5cm}|}
\toprule
 \textbf{Approach(es)}& \textbf{Scope} & \textbf{Protocol(s)} & 
 \textbf{Virtualization} & \textbf{Scalability}\\[2ex]
\hline
MobIoTSim\cite{Kertesz:19}\cite{Pflanzner:16}, AWS IoT Device Simulator\cite{AmazonIoT} & IoT device& MQTT&$\times$& $\times$\\
\hline
NuvIot\cite{NuvIoT}&IoT device& REST, MQTT, AMQP, TCP, UDP & $\times$ & $\times$\\
\hline
IoTSim-Edge\cite{Jha:20}& IoT device, edge device & MQTT, AMQP, XMPP, CoAP& $\times$ & $\times$\\
\hline
IoTCloudSamples\cite{Truong:21} & IoT device, edge device &MQTT, AMQP, REST, CoAP& $\checkmark$ & $\times$
\\
\hline
EdgeCloudSim~\cite{Sonmez:17} & IoT device, edge device, network, cloud & Unspecified & $\checkmark$ & $\times$\\
\hline
OMNet++\cite{OMNeT++}, NS-3\cite{NS-3}, Castalia\cite{boulis:11}& network & MQTT, CoAP, TCP, UDP, etc.& $\times$ & $\checkmark$\\
\hline
COOJA\cite{Osterlind:06}, Mininet\cite{Mininet}, MaxiNet\cite{Wette:14}& network & MQTT, CoAP, TCP, UDP, etc. & $\times$ & $\times$\\
\hline
PIoT\cite{Firouzabadi:23}&IoT device, network&Unspecified&$\times$&$\times$\\
\hline
Kaala\cite{Dayalan:22}&IoT device, network, cloud &MQTT, CoAP, TCP, UDP, etc. & $\checkmark$ & $\times$\\
\hline
IOTSim\cite{Zeng:17}, GreenCloud\cite{Kliazovich:12}, CloudSim\cite{Calheiros:2009}, NetworkCloudSim\cite{Garg:11}&cloud&Unspecified&$\times$&$\times$\\
\hline
Moawad et al.\cite{Moawad:15}&data&Unspecified&$\times$&$\times$\\
\hline
EMU-IoT\cite{Ramprasad:19}&IoT device, edge device, cloud&MQTT, HTTP&$\checkmark$&$\times$\\
\hline
IOTier\cite{Nikolaidis:21}&IoT device, edge device, network&4G, Starlink, 802.11n, DSL, etc&$\checkmark$&$\times$\\
\hline
Fogify\cite{Symeonides:20}&edge device&VXLAN&$\checkmark$&$\times^*$\\
\hline
UiTiOt v3\cite{Ly-Trong:18}&IoT device&802.11a/b/g, ZigBee&$\checkmark$&$\times^*$\\
\hline
Sendorek et al.\cite{Sendorek:18}&IoT device, edge device&MQTT, GPIO, I2C, etc&$\checkmark$&$\times$\\
\hline
iFogSim\cite{Gupta:17}&IoT device, edge device& Unspecified&$\checkmark$&$\times$\\
\hline
FogTorch\cite{Brogi:17}&edge device&Unspecified &$\times$&$\times$\\
\hline
FogNetSim++\cite{Qayyum:18}&IoT device, edge device, network& TCP, UDP, FTP, HTTP, MQTT, CoAP, AMPQ&$\times$&$\times$\\
\hline
Fogbed\cite{Coutinho:18}&edge device, fog device, network, cloud&Unspecified&$\checkmark$&$\times$\\
\hline
EmuFog\cite{Mayer:17}&fog device, network&Unspecified&$\checkmark$&$\times$\\
\hline
\lang&edge device, cloud &MQTT, TCP, UDP&$\checkmark$&$\checkmark$\\
\bottomrule
\multicolumn{5}{l}{\raggedright\footnotesize{\it * These approaches provide partial support for scalability but cannot go beyond limits of physical computational resources.}}
\end{tabular}
}
\end{center}

\end{table*}

To better position our work within the  literature on IoT simulation and testing,  we present in Table~\ref{tab:relatedwork} a structured comparison with the research strands and online tools most closely related to our work. We use the following criteria for the comparison:

(1)~\emph{Scope} refers to the IoT layer(s) that  an approach simulates or addresses in terms of the particular testing challenges of the layer(s). Common IoT layers addressed by IoT simulators include  the following:  \emph{device}, \emph{edge}, \emph{fog}, \emph{network}, and \emph{cloud}. Each of  these layers possesses specific characteristics and details, and entails different analytical needs. Due to the heterogeneity and intrinsic differences among these layers, IoT simulators  and testbeds usually focus on just one or two of these layers. Since fog and edge layers often serve similar functions, they are sometimes used interchangeably in the IoT literature~\cite{Chiang:16}. In our classification of IoT simulators and testbeds, we consider the terms "fog" and "edge" to be synonymous unless a clear distinction between fog and edge is made by the simulator or the testbed. In addition, some methods are focused exclusively on the analysis and synthesis of the data generated and transmitted between different layers, and do not explicitly capture any IoT layer. In Table~\ref{tab:relatedwork}, the focus of these methods is denoted as ``data''. 

(2)~\emph{Protocol(s)} indicate the network protocols that an approach supports.  As indicated in Table~\ref{tab:relatedwork}, IoT simulators that are focused on the network layer, e.g.,~\cite{OMNeT++,Cooja,NS-3,Mininet,boulis:11}, may support several protocols.  IoT simulators and testbeds, which are focused on evaluating  edge / fog computing without specifying a network protocol, are denoted ``Unspecified'' in Table~\ref{tab:relatedwork}. 

(3)~\emph{Virtualization} indicates whether a simulation or testing approach uses any virtualization techniques, such as virtual machines and containers, to enhance its effectiveness and efficiency.

(4)~\emph{Scalability} denotes whether a simulation or testing approach  has mechanisms to expand its operational capacity when exceeding the computational resource limits. In other words, this criterion determines if  a simulator or a testbed offers configurable parameters or  other methods for scaling up its capacity. Depending on the focus, scaling up the capacity may involve  increasing the number of simulated IoT or edge devices, enlarging the simulated network, or raising the number of microservices in a cloud system.

Having set the stage with Table~\ref{tab:relatedwork}, we now discuss the specific work strands listed in the table. We begin with an overview of IoT simulations, followed by an outline of testbeds and techniques designed for testing IoT systems. We then contrast our work with existing IoT simulation and testing approaches.

\textbf{IoT Simulation.} Numerous simulators exist for different components of IoT systems, e.g., IoT sensors and actuators~\cite{Kertesz:19,AmazonIoT,Dayalan:22,Pflanzner:16,NuvIoT}, IoT edge devices~\cite{Jha:20,Truong:21,Sonmez:17}, IoT networks~\cite{OMNeT++,Osterlind:06,NS-3, Mininet,boulis:11,Wette:14}, and IoT cloud infrastructures~\cite{Calheiros:2009,Garg:11,Zeng:17,Kliazovich:12}.
Among these, the closest ones to our work are IoT edge simulators. 

IoTCloudSamples~\cite{Truong:21} 
provides a simluation framework aimed at facilitating the development and operation of edge software systems. IoTSim-Edge~\cite{Jha:20} is an IoT device-to-edge simulator enabling the specification of various IoT-device characteristics, e.g., network connectivity, mobility and energy consumption, and simulating the interactions of IoT devices with edge devices. 
EdgeCloudSim~\cite{Sonmez:17} is an edge computing simulator providing features such as network modelling, edge-server simulation, and packet generation  for IoT devices. iFogSim~\cite{Gupta:17} is an IoT device-to-edge simulator focusing on edge computing. It is used to evaluate resource management and application scheduling policies across edge devices. FogTorch~\cite{Brogi:17} is a model designed for determining QoS-aware deployments of IoT applications in fog infrastructures. It accepts an infrastructure and an application as inputs and employs a heuristic search approach to identify suitable deployments on edge devices.  FogNetSim++~\cite{Qayyum:18} is a simulator designed for task execution in the fog or edge layer, featuring a centralized broker that facilitates task scheduling and mobility-driven handover for IoT devices. In addition to its system model, FogNetSim++ offers an energy model and a pricing model, enabling users to explore various request scheduling and resource handover algorithms in fog computing environments. Fogbed~\cite{Coutinho:18} enhances Mininet~\cite{Mininet}, a network simulator,  by adding support for Docker containers as virtual nodes within its edge and fog layers. 
Combining Mininet's functions with user-provided Docker images for edge, fog, and cloud, FogBed enables the seamless integration of various computing elements, helping users in designing and testing fog applications. Emufog~\cite{Mayer:17} is an emulation framework based on MaxiNet~\cite{Wette:14}, a network simulator. This framework is designed for creating network topologies, placing fog nodes within these topologies, and executing applications on the fog nodes. Emufog further provides a latency-based fog node placement policy, facilitating the selection of simulated routers for ideal fog node deployment.

\textbf{IoT Testing.} Testing IoT systems is a challenging task due to numerous factors, e.g., the sheer scale of IoT systems, the heterogeneity of IoT sensors and actuators, the diversity of network protocols and network topologies in IoT systems, and the tight integration of IoT systems with their environment~\cite{Beilharz:2021}.

Motivated by testing IoT systems, Moawad et al.~\cite{Moawad:15} propose a conceptualization of IoT data at run-time. They then use time-series compression and polynomial segmentation algorithms 
to synthesize realistic sensor data based on their proposed conceptualization.

EMU-IoT~\cite{Ramprasad:19} is an IoT testbed based on a microservice architecture. This testbed can emulate IoT devices through containers running on virtual machines, create virtualized gateways, and define orchestrators, monitors and load balancers. EMU-IoT has been applied for various testing purposes, e.g., testing the collection of information about resource consumption across an IoT system.
IOTier~\cite{Nikolaidis:21} is a testbed that enables using resource-constrained containers as IoT components and grouping these components into a device tier, gateway tier and cloud tier. Using IOTier, one can simulate dynamic changes in the operating conditions of IoT systems and thereby test device capabilities, network performance and load-balancing strategies. 
Fogify\cite{Symeonides:20} is an edge testbed that focuses on emulating edge topologies and deployment conditions using containerized descriptions. Fogify further provides a mechanism to alter network quality and inject entity and infrastructure downtime at run-time. 
UiTiOt (version 3)~\cite{Ly-Trong:18} provides a testbed for integrating real IoT devices with emulated devices that run as containers on virtual machines to support large-scale experimentation.
Sendorek et al.~\cite{Sendorek:18} propose the concept of a software-defined IoT test environment. 
Their proposed testbed can define virtual environments, virtual sensors and virtual communication devices as software, thereby allowing the combination of virtual and real equipment for testing IoT applications and communication protocols.

\textbf{Comparison.}  
Most IoT simulators and testbeds in Table~\ref{tab:relatedwork} are not  edge-to-cloud because they either concentrate mainly on the information flow between IoT devices and the edge, or they exclusively target the network or cloud layers. Only two simulators, Fogbed and EdgeCloudSim, along with one testbed, EMU-IoT, are edge-to-cloud. Among these three, EdgeCloudSim~\cite{Sonmez:17} focuses on evaluating the performance of different edge computing architectures, rather than assessing the performance of cloud systems as done by \lang. In EdgeCloudSim, data packets are communicated primarily between IoT devices and edge devices, with some packets from IoT devices directed to the cloud when the architecture includes a cloud system. Due to the differences in focus and function  between EdgeCloudSim and \lang, we did not select EdgeCloudSim for a detailed design comparison in Section~\ref{sec:usecase}. On the other hand, EMU-IoT and Fogbed are comparable to \lang\ in terms of their function and are included in our detailed comparison in Section~\ref{sec:usecase}.

Depending on specific application scenarios, IoT simulation and testing approaches  support different network communication protocols. Nevertheless, MQTT stands out as the most popular choice for IoT systems.
Our approach, \lang, which focuses on simulating edge devices for stress testing cloud systems, supports MQTT, TCP, and UDP since these protocols are widely used in IoT systems.

Some existing  IoT simulators and testbeds ($11$ out of $29$) use virtualization techniques to improve the fidelity of the simulated devices or to efficiently manage simulated entities. However,  none of these simulators or testbeds support the objective that motivates our work, i.e., stress testing of cloud systems via a simulated edge layer. Furthermore, none of the approaches that use virtualization and containerization methods perform systematic experiments that compare containers versus virtual machines and native hosts for simulation. Hence, in addition to being novel in terms of the objective and focus, our evaluation is novel in its empirically grounded examination of alternative  platforms  for edge-device simulation.

The network simulators,   OMNet++, NS-3, and Castalia~\cite{OMNeT++,NS-3, boulis:11}, support scalability by  providing mechanisms such as  parallel and distributed simulation, along with dynamic allocation and deallocation of resources during simulation. These simulators are, however, only focused on the network layer. They neither tackle stress testing for IoT cloud systems nor offer support for capturing the IoT edge and cloud layers as done in our work. Some IoT simulation and testing techniques provide partial support for  scalability such as  Fogify \cite{Symeonides:20} and UiTiOt v3~\cite{Ly-Trong:18}. Specifically, Fogify \cite{Symeonides:20} allows for the addition and removal of simulated nodes or entities,  while  UiTiOt v3~\cite{Ly-Trong:18} can increase the number of simulated IoT devices using virtual machines. However, such scalability supports are not effective as the number of simulated devices and entities is restricted by the  available physical computational resources. In contrast, \lang\ offers the flexibility to specify offset ranges and speed values. By increasing the offset range or decreasing the speed, users can increase the number of simulated IoT and edge devices, thereby enhancing the simulator's scalability  without requiring additional physical resources.

\subsection{Domain-Specific Languages (DSLs) for IoT}

Model-driven engineering (MDE) is a well-established thrust in IoT~\cite{Morin:17,Song:20,Moawad:15}. 
DSLs are particularly common as a way to abstract away from the complexity of IoT systems and thereby simplify the construction and analysis of this type of systems.

Sneps{-}Sneppe and Namiot~\cite{Sneppe:15} propose a web-based DSL to simplify data access in IoT web applications. This DSL supports both synchronous and asynchronous data updates as well as communication between processes and sensors.
Tichy et al.~\cite{Tichy:20} present a DSL that supports the declarative specification and execution of IoT system components for quality assurance purposes.
Gomes~et~al.~\cite{Gomes:17} propose EL4IoT, a DSL that aims to simplify the development of IoT-device applications by refactoring the network stack.
Amrani et al.~\cite{Amrani:17} propose IoTDSL for specifying and assembling usage scenarios of IoT systems in a way that would be understandable for end-users without programming expertise. {\AA}kesson et al.~\cite{Akesson:19} develop a DSL for IoT service composition. This DSL  supports live programming, drag-and-drop, and sequential, alternative and parallel events. Finally, Cacciagrano and Culmone~\cite{Cacciagrano:20} propose a DSL, named IRON, for programming smart environments based on Event-Condition-Action (ECA) rules~\cite{Cacciagrano:18}. The end goal here is to provide the ability to intercept specific program anomalies and \hbox{application-specific safety issues.}

These existing DSLs have helped us better shape our approach; nevertheless, none of these DSLs were designed to tackle our use case of interest, namely, stress testing of cloud systems. Specifically, the flexibility we provide for using different simulation platforms without requiring the users to have a deep understanding of virtualization and containerization technologies, and the mechanisms our DSL offers for scaling up the number of edge devices are, to the best of our knowledge, novel.

\section{Conclusion}
\label{sec:con}

In this article, we presented a lean simulation framework specifically designed for IoT cloud stress testing. This framework enables  efficient simulation of large arrays of IoT and edge devices interacting with the cloud. To simplify simulation construction  for practitioners, we proposed a domain-specific language, named  \lang, tailored for generating simulators from high-level model-based specifications. The language has been implemented using Xtext and Xtend and is publicly available~\cite{IoTECS}.

\lang\ is the result of a collaborative effort with an industry partner, Cheetah Networks. The language has been used independently by our partner in various testing and demonstration campaigns. Based on feedback from engineers at our partner, \lang\ is valuable for stress testing IoT cloud systems, saving them substantial amounts of effort.
 
 Our empirical evaluation revolved around two industrial case studies: one related to IoT connected vehicles and the other to the monitoring and generation of analytics for IoT networks. We systematically examined different platforms for running \lang\ simulators, and concluded that peak performance is achieved by containerizing groups of edge devices with Dockers, rather than using virtual machines or running (simulated) edge devices directly on host machines. Subsequently, we assessed the service capacity of our industrial case studies  using \lang\ simulators, and demonstrated that \lang\ simulators can handle 3.5 times more IoT and edge devices compared to two industrial stress-testing baseline tools, JMeter and Locust.

As explained in Section~\ref{subsec:limitation}, our approach has some limitations. While these limitations have not hindered adoption by our industry partner, addressing them in future work can further improve the usefulness and applicability of \lang. Notably, our future work aims to enable dynamic adjustments through high-level constructs for more flexible configuration and to enhance \lang\ by providing real-time access to metrics during stress testing. 
Another avenue for future work is to make simulations cost-aware. IoT solution providers often use externally hosted resources for stress testing, incurring fees based on CPU and memory usage for containers and virtual machines. Accounting for cost factors, particularly hosting expenses, is therefore important for simulation DSLs, as it offers a means to reduce simulation costs. We plan to study optimization techniques, including search algorithms, to find the optimal balance between the total number of simulated IoT devices and simulation costs.

\section{Implementation and Availability} 
We have implemented \lang\  using Xtext version 2.25.0~\cite{Xtext} and Xtend~version 2.25.0~\cite{Xtend}. The \lang\ grammar and our code generation tool are publicly available~\cite{IoTECS}. Our online material include: (1)~a complete description and implementation of \lang\ including an example of \lang\ specification, (2)~scripts for running baseline tools (i.e., JMeter and Locust), and (3)~the results of our experiments.

\section*{Acknowledgment}
We gratefully acknowledge funding from Mitacs Accelerate, Cheetah Networks, Stratosfy, Ontario Centre of Innovation, and NSERC of Canada under the Discovery and Discovery Accelerator programs.

\bibliographystyle{IEEEtran}

\begin{thebibliography}{10}
\providecommand{\url}[1]{#1}
\csname url@samestyle\endcsname
\providecommand{\newblock}{\relax}
\providecommand{\bibinfo}[2]{#2}
\providecommand{\BIBentrySTDinterwordspacing}{\spaceskip=0pt\relax}
\providecommand{\BIBentryALTinterwordstretchfactor}{4}
\providecommand{\BIBentryALTinterwordspacing}{\spaceskip=\fontdimen2\font plus
\BIBentryALTinterwordstretchfactor\fontdimen3\font minus \fontdimen4\font\relax}
\providecommand{\BIBforeignlanguage}[2]{{%
\expandafter\ifx\csname l@#1\endcsname\relax
\typeout{** WARNING: IEEEtran.bst: No hyphenation pattern has been}%
\typeout{** loaded for the language `#1'. Using the pattern for}%
\typeout{** the default language instead.}%
\else
\language=\csname l@#1\endcsname
\fi
#2}}
\providecommand{\BIBdecl}{\relax}
\BIBdecl

\bibitem{IoT17}
M.~Hung, ``Leading the iot,'' Documentation at \url{https://www.gartner.com/imagesrv/books/iot/iotEbook_digital.pdf}, 2017.

\bibitem{Chan:04}
H.~Chan, ``Accelerated stress testing for both hardware and software,'' in \emph{Annual Symposium Reliability and Maintainability, 2004 - RAMS}, 2004, pp. 346--351.

\bibitem{Shin:20}
S.~Y. Shin, S.~Nejati, M.~Sabetzadeh, L.~C. Briand, C.~Arora, and F.~Zimmer, ``Dynamic adaptation of software-defined networks for iot systems: a search-based approach,'' in \emph{{SEAMS} '20: {IEEE/ACM} 15th International Symposium on Software Engineering for Adaptive and Self-Managing Systems, Seoul, Republic of Korea, 29 June - 3 July, 2020}, 2020, pp. 137--148.

\bibitem{Li:22}
J.~Li, S.~Nejati, and M.~Sabetzadeh, ``Learning self-adaptations for iot networks: A genetic programming approach,'' in \emph{{SEAMS} '22: {IEEE/ACM} 17th International Symposium on Software Engineering for Adaptive and Self-Managing Systems, Pittsburgh, PA, USA, May 21 - 29, 2022}, 2022.

\bibitem{Li:23}
------, ``Using genetic programming to build self-adaptivity into software-defined networks,'' \emph{ACM Trans. Auton. Adapt. Syst.}, aug 2023.

\bibitem{Kertesz:19}
A.~Kert{\'{e}}sz, T.~Pflanzner, and T.~Gyim{\'{o}}thy, ``A mobile iot device simulator for iot-fog-cloud systems,'' \emph{Journal of Grid Computing}, vol.~17, no.~3, pp. 529--551, 2019.

\bibitem{AmazonIoT}
Amazon, ``Amazon web services. iot core.'' \url{https://aws.amazon.com/solutions/\\implementations/iot-device-simulator/}, 2021.

\bibitem{NuvIoT}
S.~Logistics, ``Nuviot iot simulator,'' \url{https://www.nuviot.com}, 2019.

\bibitem{Jha:20}
D.~N. Jha, K.~Alwasel, A.~Alshoshan, X.~Huang, R.~K. Naha, S.~K. Battula, S.~Garg, D.~Puthal, P.~James, A.~Y. Zomaya, S.~Dustdar, and R.~Ranjan, ``Iotsim-edge: {A} simulation framework for modeling the behavior of internet of things and edge computing environments,'' \emph{Software - Practice and Experience}, vol.~50, pp. 844--867, 2020.

\bibitem{Sonmez:17}
C.~Sonmez, A.~Ozgovde, and C.~Ersoy, ``Edgecloudsim: An environment for performance evaluation of edge computing systems,'' in \emph{Second International Conference on Fog and Mobile Edge Computing, {FMEC} 2017, Valencia, Spain, May 8-11, 2017}, 2017, pp. 39--44.

\bibitem{Ramprasad:19}
B.~Ramprasad, M.~Fokaefs, J.~Mukherjee, and M.~Litoiu, ``Emu-iot - {A} virtual internet of things lab,'' in \emph{2019 {IEEE} International Conference on Autonomic Computing, {ICAC} 2019, Ume{\aa}, Sweden, June 16-20, 2019}, 2019, pp. 73--83.

\bibitem{Kliazovich:12}
D.~Kliazovich, P.~Bouvry, and S.~U. Khan, ``Greencloud: a packet-level simulator of energy-aware cloud computing data centers,'' \emph{The Journal of Supercomputing}, vol.~62, no.~3, pp. 1263--1283, 2012.

\bibitem{Osterlind:06}
F.~{\"{O}}sterlind, A.~Dunkels, J.~Eriksson, N.~Finne, and T.~Voigt, ``Cross-level sensor network simulation with {COOJA},'' in \emph{{LCN} 2006, The 31st Annual {IEEE} Conference on Local Computer Networks, Tampa, Florida, USA, 14-16 November 2006}, 2006, pp. 641--648.

\bibitem{boulis:11}
A.~Boulis, ``Castalia: A simulator for wireless sensor networks and body area networks,'' \emph{NICTA: National ICT Australia}, vol.~83, p.~7, 2011.

\bibitem{Mininet}
Mininet, ``Mininet,'' \url{https://mininet.org}, 2021.

\bibitem{Elazhary:19}
H.~Elazhary, ``Internet of things (iot), mobile cloud, cloudlet, mobile iot, iot cloud, fog, mobile edge, and edge emerging computing paradigms: Disambiguation and research directions,'' \emph{Journal of Network and Computer Applications}, vol. 128, pp. 105--140, 2019.

\bibitem{Pflanzner:16}
T.~Pflanzner, A.~Kert{\'{e}}sz, B.~Spinnewyn, and S.~Latr{\'{e}}, ``Mobiotsim: Towards a mobile iot device simulator,'' in \emph{4th {IEEE} International Conference on Future Internet of Things and Cloud Workshops, FiCloud Workshops 2016, Vienna, Austria, August 22-24, 2016}, 2016, pp. 21--27.

\bibitem{Li:22-2}
J.~Li, S.~Nejati, M.~Sabetzadeh, and M.~McCallen, ``A domain-specific language for simulation-based testing of iot edge-to-cloud solutions,'' in \emph{Proceedings of the 25th International Conference on Model Driven Engineering Languages and Systems, {MODELS} 2022, Montreal, Quebec, Canada, October 23-28, 2022}, 2022, pp. 367--378.

\bibitem{Coutinho:18}
A.~Coutinho, F.~Greve, C.~V.~S. Prazeres, and J.~Cardoso, ``Fogbed: {A} rapid-prototyping emulation environment for fog computing,'' in \emph{2018 {IEEE} International Conference on Communications, {ICC} 2018, Kansas City, MO, USA, May 20-24, 2018}.\hskip 1em plus 0.5em minus 0.4em\relax {IEEE}, 2018, pp. 1--7.

\bibitem{jmeter}
\BIBentryALTinterwordspacing
{Apache JMeter}, The Apache Software Foundation, 2023. [Online]. Available: \url{https://jmeter.apache.org/}
\BIBentrySTDinterwordspacing

\bibitem{locust}
{Locust}, \url{https://locust.io}, 2023.

\bibitem{IoTECS}
SednaLab, ``Iotecs,'' \url{https://github.com/sednalab/IoTECS/tree/main}, 2023.

\bibitem{Pahl:19}
C.~Pahl, A.~Brogi, J.~Soldani, and P.~Jamshidi, ``Cloud container technologies: {A} state-of-the-art review,'' \emph{{IEEE} Trans. Cloud Comput.}, vol.~7, no.~3, pp. 677--692, 2019.

\bibitem{Dizdarevic:19}
J.~Dizdarevic, F.~Carpio, A.~Jukan, and X.~Masip{-}Bruin, ``A survey of communication protocols for internet of things and related challenges of fog and cloud computing integration,'' \emph{{ACM} Comput. Surv.}, vol.~51, no.~6, pp. 116:1--116:29, 2019.

\bibitem{Aksakalli:21}
I.~K. Aksakalli, T.~{\c{C}}elik, A.~B. Can, and B.~Tekinerdogan, ``Deployment and communication patterns in microservice architectures: {A} systematic literature review,'' \emph{J. Syst. Softw.}, vol. 180, p. 111014, 2021.

\bibitem{Varghese:16}
B.~Varghese, N.~Wang, S.~Barbhuiya, P.~Kilpatrick, and D.~S. Nikolopoulos, ``Challenges and opportunities in edge computing,'' in \emph{2016 {IEEE} International Conference on Smart Cloud, SmartCloud 2016, New York, NY, USA, November 18-20, 2016}.\hskip 1em plus 0.5em minus 0.4em\relax {IEEE} Computer Society, 2016, pp. 20--26.

\bibitem{Mani:18}
S.~K. Mani, R.~Durairajan, P.~Barford, and J.~Sommers, ``An architecture for iot clock synchronization,'' in \emph{Proceedings of the 8th International Conference on the Internet of Things, {IOT} 2018, Santa Barbara, CA, USA, October 15-18, 2018}, 2018, pp. 17:1--17:8.

\bibitem{Scheffer:95}
M.~Scheffer, J.~Baveco, D.~DeAngelis, K.~Rose, and E.~{van Nes}, ``Super-individuals a simple solution for modelling large populations on an individual basis,'' \emph{Ecological Modelling}, vol.~80, no.~2, pp. 161--170, 1995.

\bibitem{Xtext}
{Eclipse Foundation, Inc.}, ``Xtext v2.25.0,'' \url{https://www.eclipse.org/Xtext/}, 2021.

\bibitem{Xtend}
------, ``Xtend v2.25.0,'' \url{https://www.eclipse.org/Xtend/}, 2021.

\bibitem{Araujo:19}
V.~Araujo, K.~Mitra, S.~Saguna, and C.~{\AA}hlund, ``Performance evaluation of {FIWARE:} {A} cloud-based iot platform for smart cities,'' \emph{J. Parallel Distributed Comput.}, vol. 132, pp. 250--261, 2019.

\bibitem{Ismail:18}
A.~A. Ismail, H.~S. Hamza, and A.~M. Kotb, ``Performance evaluation of open source iot platforms,'' in \emph{2018 IEEE Global Conference on Internet of Things (GCIoT)}, 2018.

\bibitem{Lazidis:22}
A.~Lazidis, K.~Tsakos, and E.~G. Petrakis, ``Publish--subscribe approaches for the iot and the cloud: Functional and performance evaluation of open-source systems,'' \emph{Internet of Things}, vol.~19, p. 100538, 2022.

\bibitem{Hou:17}
L.~Hou, S.~Zhao, X.~Li, P.~Chatzimisios, and K.~Zheng, ``Design and implementation of application programming interface for internet of things cloud,'' \emph{International Journal of Network Management}, vol.~27, no.~3, p. e1936, 2017.

\bibitem{Koziolek:20}
H.~Koziolek, S.~Gr{\"{u}}ner, and J.~R{\"{u}}ckert, ``A comparison of {MQTT} brokers for distributed iot edge computing,'' in \emph{Software Architecture - 14th European Conference, {ECSA} 2020, L'Aquila, Italy, September 14-18, 2020, Proceedings}, ser. Lecture Notes in Computer Science, vol. 12292, 2020, pp. 352--368.

\bibitem{Pajooh:21}
H.~Honar~Pajooh, M.~A. Rashid, F.~Alam, and S.~Demidenko, ``Iot big data provenance scheme using blockchain on hadoop ecosystem,'' \emph{Journal of Big Data}, vol.~8, no.~1, p. 114, 2021.

\bibitem{kubernetes}
{Kubernetes}, \url{https://kubernetes.io}, 2023.

\bibitem{lambdaArchitecture}
N.~Marz and J.~Warren, \emph{{Big Data: Principles and Best Practices of Scalable Real-Time Data Systems}}.\hskip 1em plus 0.5em minus 0.4em\relax Manning Publications Co., 2015.

\bibitem{jmeterplugins}
A.~Pokhilko, ``{UDPRequest},'' \url{https://jmeter-plugins.org/wiki/UDPRequest/}, 2023.

\bibitem{Carmona:22}
O.~G{\'{o}}mez{-}Carmona, D.~Casado{-}Mansilla, D.~L{\'{o}}pez{-}de{-}Ipi{\~{n}}a, and J.~Garc{\'{\i}}a{-}Zub{\'{\i}}a, ``Optimizing computational resources for edge intelligence through model cascade strategies,'' \emph{{IEEE} Internet Things J.}, vol.~9, no.~10, pp. 7404--7417, 2022.

\bibitem{Hoh:07}
B.~Hoh, M.~Gruteser, H.~Xiong, and A.~Alrabady, ``Preserving privacy in gps traces via uncertainty-aware path cloaking,'' in \emph{Proceedings of the 2007 {ACM} Conference on Computer and Communications Security, {CCS} 2007, Alexandria, Virginia, USA, October 28-31, 2007}.\hskip 1em plus 0.5em minus 0.4em\relax {ACM}, 2007, pp. 161--171.

\bibitem{Sonklin:19}
K.~Sonklin, C.~Wang, D.~Jayalath, and Y.~Feng, ``Connected-vehicle data exchanges and positioning computing based on the publish-subscribe paradigm,'' \emph{Journal of Computer and Communications}, vol.~07, pp. 82--93, 01 2019.

\bibitem{Wireshark}
{Wireshark Foundation}, ``Wireshark v3.6.0,'' \url{https://www.wireshark.org/}, 2021.

\bibitem{iPerf}
J.~Dugan, S.~Elliott, B.~A. Mah, J.~Poskanzer, and K.~Prabhu, ``iperf,'' \url{https://iperf.fr}, 2024.

\bibitem{soni:17}
D.~Soni and A.~Makwana, ``A survey on mqtt: a protocol of internet of things (iot),'' in \emph{International conference on telecommunication, power analysis and computing techniques (ICTPACT-2017)}, vol.~20, 2017, pp. 173--177.

\bibitem{kumar:12}
S.~Kumar and S.~Rai, ``Survey on transport layer protocols: Tcp \& udp,'' \emph{International Journal of Computer Applications}, vol.~46, no.~7, pp. 20--25, 2012.

\bibitem{Truong:21}
H.~L. Truong, ``Using iotcloudsamples as a software framework for simulations of edge computing scenarios,'' \emph{Internet Things}, vol.~14, p. 100383, 2021.

\bibitem{OMNeT++}
{OMNet++}, ``{OMNeT++},'' \url{https://omnetpp.org/intro/}, 2022.

\bibitem{NS-3}
NS-3, ``Ns-3.35,'' \url{https://www.nsnam.org}, 2021.

\bibitem{Wette:14}
P.~Wette, M.~Dr{\"{a}}xler, and A.~Schwabe, ``Maxinet: Distributed emulation of software-defined networks,'' in \emph{2014 {IFIP} Networking Conference, Trondheim, Norway, June 2-4, 2014}.\hskip 1em plus 0.5em minus 0.4em\relax {IEEE} Computer Society, 2014, pp. 1--9.

\bibitem{Firouzabadi:23}
A.~D. Firouzabadi, H.~Mellah, O.~Manzanilla{-}Salazar, R.~Khalvandi, V.~Therrien, V.~Boutin, and B.~Sans{\`{o}}, ``Piot: {A} performance iot simulation system for a large-scale city-wide assessment,'' \emph{{IEEE} Access}, vol.~11, pp. 56\,273--56\,286, 2023.

\bibitem{Dayalan:22}
U.~K. Dayalan, R.~A.~K. Fezeu, T.~J. Salo, and Z.~Zhang, ``Kaala: scalable, end-to-end, iot system simulator,'' in \emph{NET4us '22: Proceedings of the {ACM} {SIGCOMM} Workshop on Networked Sensing Systems for a Sustainable Society, Amsterdam, The Netherlands, August 22, 2022}.\hskip 1em plus 0.5em minus 0.4em\relax {ACM}, 2022, pp. 33--38.

\bibitem{Zeng:17}
X.~Zeng, S.~K. Garg, P.~Strazdins, P.~P. Jayaraman, D.~Georgakopoulos, and R.~Ranjan, ``Iotsim: {A} simulator for analysing iot applications,'' \emph{Journal of Systems Architecture: Embedded Software Design}, vol.~72, pp. 93--107, 2017.

\bibitem{Calheiros:2009}
R.~N. Calheiros, R.~Ranjan, C.~A. F.~D. Rose, and R.~Buyya, ``Cloudsim: {A} novel framework for modeling and simulation of cloud computing infrastructures and services,'' \emph{CoRR}, vol. abs/0903.2525, 2009.

\bibitem{Garg:11}
S.~K. Garg and R.~Buyya, ``Networkcloudsim: Modelling parallel applications in cloud simulations,'' in \emph{{IEEE} 4th International Conference on Utility and Cloud Computing, {UCC} 2011, Melbourne, Australia, December 5-8, 2011}, 2011, pp. 105--113.

\bibitem{Moawad:15}
A.~Moawad, T.~Hartmann, F.~Fouquet, G.~Nain, J.~Klein, and Y.~L. Traon, ``Beyond discrete modeling: {A} continuous and efficient model for iot,'' in \emph{18th {ACM/IEEE} International Conference on Model Driven Engineering Languages and Systems, MoDELS 2015, Ottawa, ON, Canada, September 30 - October 2, 2015}, 2015, pp. 90--99.

\bibitem{Nikolaidis:21}
F.~Nikolaidis, M.~Marazakis, and A.~Bilas, ``Iotier: {A} virtual testbed to evaluate systems for iot environments,'' in \emph{21st {IEEE/ACM} International Symposium on Cluster, Cloud and Internet Computing, CCGrid 2021, Melbourne, Australia, May 10-13, 2021}, 2021, pp. 676--683.

\bibitem{Symeonides:20}
M.~Symeonides, Z.~Georgiou, D.~Trihinas, G.~Pallis, and M.~D. Dikaiakos, ``Fogify: {A} fog computing emulation framework,'' in \emph{5th {IEEE/ACM} Symposium on Edge Computing, {SEC} 2020, San Jose, CA, USA, November 12-14, 2020}, 2020, pp. 42--54.

\bibitem{Ly-Trong:18}
N.~Ly{-}Trong, C.~Dang{-}Le{-}Bao, D.~Huynh{-}Van, and Q.~L. Trung, ``Uitiot v3: {A} hybrid testbed for evaluation of large-scale iot networks,'' in \emph{Proceedings of the Ninth International Symposium on Information and Communication Technology, SoICT 2018, Danang City, Vietnam, December 06-07, 2018}, 2018, pp. 155--162.

\bibitem{Sendorek:18}
J.~Sendorek, T.~Szydlo, and R.~Brzoza{-}Woch, ``Software-defined virtual testbed for iot systems,'' \emph{Wireless Communications and Mobile Computing}, vol. 2018, pp. 1\,068\,261:1--1\,068\,261:11, 2018.

\bibitem{Gupta:17}
H.~Gupta, A.~V. Dastjerdi, S.~K. Ghosh, and R.~Buyya, ``ifogsim: {A} toolkit for modeling and simulation of resource management techniques in the internet of things, edge and fog computing environments,'' \emph{Softw. Pract. Exp.}, vol.~47, no.~9, pp. 1275--1296, 2017.

\bibitem{Brogi:17}
A.~Brogi and S.~Forti, ``Qos-aware deployment of iot applications through the fog,'' \emph{{IEEE} Internet Things J.}, vol.~4, no.~5, pp. 1185--1192, 2017.

\bibitem{Qayyum:18}
T.~Qayyum, A.~Malik, M.~Khattak, O.~Khalid, and S.~Khan, ``Fognetsim++: A toolkit for modeling and simulation of distributed fog environment,'' \emph{IEEE Access}, vol.~PP, pp. 1--1, 10 2018.

\bibitem{Mayer:17}
R.~Mayer, L.~Graser, H.~Gupta, E.~Saurez, and U.~Ramachandran, ``Emufog: Extensible and scalable emulation of large-scale fog computing infrastructures,'' in \emph{{IEEE} Fog World Congress, {FWC} 2017, Santa Clara, CA, USA, October 30 - Nov. 1, 2017}.\hskip 1em plus 0.5em minus 0.4em\relax {IEEE}, 2017, pp. 1--6.

\bibitem{Chiang:16}
M.~Chiang and T.~Zhang, ``Fog and iot: An overview of research opportunities,'' \emph{{IEEE} Internet Things J.}, vol.~3, no.~6, pp. 854--864, 2016.

\bibitem{Cooja}
Cooja, ``Cooja simulator,'' \url{https://anrg.usc.edu/contiki/index.php/Cooja_Simulator}, 2016.

\bibitem{Beilharz:2021}
J.~Beilharz, P.~Wiesner, A.~Boockmeyer, L.~Pirl, D.~Friedenberger, F.~Brokhausen, I.~Behnke, A.~Polze, and L.~Thamsen, ``Continuously testing distributed iot systems: An overview of the state of the art,'' \emph{CoRR}, vol. abs/2112.09580, 2021.

\bibitem{Morin:17}
B.~Morin, N.~Harrand, and F.~Fleurey, ``Model-based software engineering to tame the iot jungle,'' \emph{{IEEE} Software}, vol.~34, no.~1, pp. 30--36, 2017.

\bibitem{Song:20}
H.~Song, R.~Dautov, N.~Ferry, A.~Solberg, and F.~Fleurey, ``Model-based fleet deployment of edge computing applications,'' in \emph{MoDELS '20: {ACM/IEEE} 23rd International Conference on Model Driven Engineering Languages and Systems, Virtual Event, Canada, 18-23 October, 2020}, 2020, pp. 132--142.

\bibitem{Sneppe:15}
M.~Sneps{-}Sneppe and D.~Namiot, ``On web-based domain-specific language for internet of things,'' \emph{CoRR}, vol. abs/1505.06713, 2015.

\bibitem{Tichy:20}
M.~Tichy, J.~Pietron, D.~M{\"{o}}dinger, K.~Juhnke, and F.~J. Hauck, ``Experiences with an internal {DSL} in the iot domain,'' in \emph{{STAF} 2020 Workshop Proceedings: 4th Workshop on Model-Driven Engineering for the Internet-of-Things, 1st International Workshop on Modeling Smart Cities, and 5th International Workshop on Open and Original Problems in Software Language Engineering co-located with Software Technologies: Applications and Foundations federation of conferences {(STAF} 2020), Bergen, Norway, June 22-26, 2020}, 2020, pp. 22--34.

\bibitem{Gomes:17}
T.~Gomes, P.~Lopes, J.~Alves, P.~Mestre, J.~Cabral, J.~L. Monteiro, and A.~Tavares, ``A modeling domain-specific language for iot-enabled operating systems,'' in \emph{{IECON} 2017 - 43rd Annual Conference of the {IEEE} Industrial Electronics Society, Beijing, China, October 29 - November 1, 2017}, 2017, pp. 3945--3950.

\bibitem{Amrani:17}
M.~Amrani, F.~Gilson, A.~Debieche, and V.~Englebert, ``Towards user-centric dsls to manage iot systems,'' in \emph{Proceedings of the 5th International Conference on Model-Driven Engineering and Software Development, {MODELSWARD} 2017, Porto, Portugal, February 19-21, 2017}, 2017, pp. 569--576.

\bibitem{Akesson:19}
A.~{\AA}kesson, G.~Hedin, M.~Nordahl, and B.~Magnusson, ``Compos: Composing oblivious services,'' in \emph{{IEEE} International Conference on Pervasive Computing and Communications Workshops, PerCom Workshops 2019, Kyoto, Japan, March 11-15, 2019}, 2019, pp. 132--138.

\bibitem{Cacciagrano:20}
D.~R. Cacciagrano and R.~Culmone, ``{IRON:} reliable domain specific language for programming iot devices,'' \emph{Internet Things}, vol.~9, p. 100020, 2020.

\bibitem{Cacciagrano:18}
D.~R. Cacciagrano, F.~Corradini, R.~Culmone, N.~Gorogiannis, L.~Mostarda, F.~Raimondi, and C.~Vannucchi, ``Analysis and verification of {ECA} rules in intelligent environments,'' \emph{Journal of Ambient Intelligence and Smart Environments}, vol.~10, no.~3, pp. 261--273, 2018.

\end{thebibliography}

\onecolumn
\begin{appendices}
\section{Interview Questionnaire and Transcripts}
\label{appendix:interviews}
\subsection{Questionnaire}
\label{appendix:questionnaire}
\vspace*{1em}\noindent\textit{\textbf{Q1.} How many years of experience do you have in the field of IoT?
}

\vspace*{1em}\noindent\textit{\textbf{Q2.} Have you been involved in verification, validation or testing of IoT cloud systems in your professional capacity? If yes, please explain.
}

\vspace*{1em}\noindent\textit{\textbf{Q3.} What challenges have you encountered in the quality assurance of IoT cloud systems? 
}

\vspace*{1em}\noindent\textit{\textbf{Q4.} What purposes are you using \lang\ for?
}

\vspace*{1em}\noindent\textit{\textbf{Q5.} Consider the purposes you are using \lang\ for (discussed under Q4). Have you used any other tools in the past for one or more of the same purposes? 
}

\vspace*{1em}\noindent\textit{\textbf{Q6.} In what ways do you perceive \lang\ to be superior or inferior to past practices?
}

\vspace*{1em}\noindent\textit{\textbf{Q7.} How much effort did you find it required to learn how to use \lang? Please try to quantify in terms of the number of hours.
}

\vspace*{1em}\noindent\textit{\textbf{Q8.} Please complete this statement:\\[.5em] I found \lang\  ($\square$Very easy, $\square$Easy, $\square$Difficult, $\square$Very difficult)
 to use.}

\vspace*{1em}\noindent\textit{\textbf{Q9.} For the projects in which you have used \lang, how much effort do you estimate \lang\ saved you compared to the best alternative tool or the best combination of other tools? Please try to quantify in person hours / days.
}

\vspace*{1em}\noindent\textit{\textbf{Q10.} What specific benefits have you observed while using \lang?
}

\vspace*{1em}\noindent\textit{\textbf{Q11.} Please rate this statement: \\[.5em]
I found \lang\ to be a useful complement to existing tools for stress testing.\\[.5em]
        \indent $\square$Strongly agree,
        $\square$Agree,
        $\square$Neutral,
        $\square$Disagree,
        $\square$Strongly disagree
}

\vspace*{1em}\noindent\textit{\textbf{Q12.} Do you have any additional comments or insights you would like to share regarding \lang? 
}

\vspace*{.3cm}

\noindent\hrulefill 
\vspace*{.4cm}

\subsection{Interview Transcript \#1}
\label{appendix:transcript1}

\vspace*{1em}\noindent\textit{\textbf{Q1.} How many years of experience do you have in the field of IoT?
}
\begin{samepage}
\begin{mdframed}[style=MyFrame]
I have 8 to 9 years of experience in IoT, with my main focus on software engineering for IoT systems.
\end{mdframed}
\end{samepage}

\vspace*{1em}\noindent\textit{\textbf{Q2.} Have you been involved in verification, validation or testing of IoT cloud systems in your professional capacity? If yes, please explain.
}
\begin{samepage}
\begin{mdframed}[style=MyFrame]
Yes, extensively.\\
I have worked on several verification and testing tasks, including verifying the quality of connections between edge devices and the cloud, checking network latency, jitter, and loss, and localizing faults to determine whether they originate from the data plane, 3G, Wifi, Ethernet, or other components of the system.
\end{mdframed}
\end{samepage}

\vspace*{1em}\noindent\textit{\textbf{Q3.} What challenges have you encountered in the quality assurance of IoT cloud systems? 
}
\begin{samepage}
\begin{mdframed}[style=MyFrame]
Finding problems, replicating them, and pinpointing troublesome areas have all been challenges in testing our IoT cloud systems. Cloud platforms are generally reliable and problems don't occur frequently. In some sense, finding issues in the cloud is like searching for a needle in a haystack. Nonetheless, we must do this because cloud reliability is crucial for both our clients and our own operations.
\end{mdframed}
\end{samepage}

\vspace*{1em}\noindent\textit{\textbf{Q4.} What purposes are you using \lang\ for?
}

\begin{samepage}
\begin{mdframed}[style=MyFrame]
Our team uses \lang\ for various purposes. Some of our developers are actually using it in the next room as we speak. One important application is generating packets within the phone to assess QoE (Quality of Experience). Another use case is testing loss intervals, jitter, and latency. A third use case is large-scale testing, where we use the tool to simulate large numbers of UEs (User Equipment), such as phones, alongside real devices.
\end{mdframed}
\end{samepage}

\vspace*{1em}\noindent\textit{\textbf{Q5.} Consider the purposes you are using \lang\ for (discussed under Q4). Have you used any other tools in the past for one or more of the same purposes? 
}

\begin{samepage}
\begin{mdframed}[style=MyFrame]
I cannot think of any professional tool that we have used for the same purposes. We have developed numerous in-house scripts and testing programs in the past for these purposes, but no particular off-the-shelf tool comes to mind.
\end{mdframed}
\end{samepage}

\vspace*{1em}\noindent\textit{\textbf{Q6.} In what ways do you perceive \lang\ to be superior or inferior to past practices?
}
\begin{samepage}
\begin{mdframed}[style=MyFrame]
I would say the main advantage of \lang\ is that it is more easily configurable for different testing scenarios compared to scripting.
\end{mdframed}
\end{samepage}

\vspace*{1em}\noindent\textit{\textbf{Q7.} How much effort did you find it required to learn how to use \lang? Please try to quantify in terms of the number of hours.
}
\begin{samepage}
\begin{mdframed}[style=MyFrame]
Minimal effort, really. I would estimate it to be under 10 minutes, assuming of course the user is familiar with cloud testing and the basics of Unix. I have also observed that the tool is easy for our other developers to pick up.
\end{mdframed}
\end{samepage}

\vspace*{1em}\noindent\textit{\textbf{Q8.} Please complete this statement:\\[.5em] I found \lang\  (\underline{\bfseries$\text{\rlap{$\checkmark$}}\square$Very easy}, $\square$Easy, $\square$Difficult, $\square$Very difficult)
 to use.}

\vspace*{1em}\noindent\textit{\textbf{Q9.} For the projects in which you have used \lang, how much effort do you estimate \lang\ saved you compared to the best alternative tool or the best combination of other tools? Please try to quantify in person hours / days.
}
\begin{samepage}
\begin{mdframed}[style=MyFrame]
As a rough estimate and thinking about three major projects in the past 6 months, I would say we have saved at least 2 dedicated person-weeks during these projects.
\end{mdframed}
\end{samepage}

\vspace*{1em}\noindent\textit{\textbf{Q10.} What specific benefits have you observed while using \lang?
}

\begin{samepage}
\begin{mdframed}[style=MyFrame]
Easy configurability stands out, as we discussed before. The main implied benefit here is higher-coverage testing. We can use \lang\ to test lots of scenarios very fast; there is no need to change code every single time. 
\end{mdframed}
\end{samepage}

\vspace*{1em}\noindent\textit{\textbf{Q11.} Please rate this statement: \\[.5em]
I found \lang\ to be a useful complement to existing tools for stress testing.\\[.5em]
        \underline{\bfseries$\text{\rlap{$\checkmark$}}\square$Strongly agree},
        $\square$Agree,
        $\square$Neutral,
        $\square$Disagree,
        $\square$Strongly disagree
}

\vspace*{1em}\noindent\textit{\textbf{Q12.} Do you have any additional comments or insights you would like to share regarding \lang? 
}
\begin{samepage}
\begin{mdframed}[style=MyFrame]
There are times when we need to introduce latencies on specific generated packets to simulate more realistic edge devices. It would be helpful to add this feature into the tool.
\end{mdframed}
\end{samepage}

\noindent\hrulefill 

\subsection{Interview Transcript \#2}
\label{appendix:transcript2}

\vspace*{1em}\noindent\textit{\textbf{Q1.} How many years of experience do you have in the field of IoT?
}
\begin{samepage}
\begin{mdframed}[style=MyFrame]
I have approximately 5 years of professional experience, mainly in network engineering and software testing.
\end{mdframed}
\end{samepage}

\vspace*{1em}\noindent\textit{\textbf{Q2.} Have you been involved in verification, validation or testing of IoT cloud systems in your professional capacity? If yes, please explain.
}
\begin{samepage}
\begin{mdframed}[style=MyFrame]
Yes.\\
I would categorize my relevant expertise as follows: (a) Full verification cycle of software, (b) Black-box testing, and (c) QA for the software cycle, primarily in IoT edge software.
\end{mdframed}
\end{samepage}

\vspace*{1em}\noindent\textit{\textbf{Q3.} What challenges have you encountered in the quality assurance of IoT cloud systems? 
}
\begin{samepage}
\begin{mdframed}[style=MyFrame]
Testing our systems for different generations of mobile communication technologies (such as LTE and 5G) and alongside new products requires adjusting testing parameters and re-developing testing tools. Automating tests during the verification cycle and the repeated changes we need to make to the testing tools, often create significant challenges.
\end{mdframed}
\end{samepage}

\vspace*{1em}\noindent\textit{\textbf{Q4.} What purposes are you using \lang\ for?
}
\begin{samepage}
\begin{mdframed}[style=MyFrame]
(a) Stress testing and (b) Network stability analysis
\end{mdframed}
\end{samepage}

\vspace*{1em}\noindent\textit{\textbf{Q5.} Consider the purposes you are using \lang\ for (discussed under Q4). Have you used any other tools in the past for one or more of the same purposes? 
}

\begin{samepage}
\begin{mdframed}[style=MyFrame]
I have been using standard tools, such as ping and iperf, for this purpose for a long time. The issue is that a lot of scripting is required to use these tools for stress testing and software validation. This effort has to be repeated not only for different systems but also for different network technologies.
\end{mdframed}
\end{samepage}

\vspace*{1em}\noindent\textit{\textbf{Q6.} In what ways do you perceive \lang\ to be superior or inferior to past practices?
}
\begin{samepage}
\begin{mdframed}[style=MyFrame]
I think \lang\ is superior to past practices in terms of scalability, specifically due to its ability to easily expand from 100 to 1,000, and even 10,000 devices and beyond. Another important advantage is \lang's straightforward configuration process, making it more user-friendly and much quicker to use compared to hand-crafted scripts.
\end{mdframed}
\end{samepage}

\vspace*{1em}\noindent\textit{\textbf{Q7.} How much effort did you find it required to learn how to use \lang? Please try to quantify in terms of the number of hours.
}
\begin{samepage}
\begin{mdframed}[style=MyFrame]
My projection is a couple of hours for a new user, including some preliminaries from Unix. The tool is well-documented. We have two developers who are currently using the tool. We could easily have them up and running with the tool.
\end{mdframed}
\end{samepage}

\vspace*{1em}\noindent\textit{\textbf{Q8.} Please complete this statement:\\[.5em] I found \lang\  ($\square$Very easy, \underline{\bfseries$\text{\rlap{$\checkmark$}}\square$Easy}, $\square$Difficult, $\square$Very difficult)
 to use.}

\vspace*{1em}\noindent\textit{\textbf{Q9.} For the projects in which you have used \lang, how much effort do you estimate \lang\ saved you compared to the best alternative tool or the best combination of other tools? Please try to quantify in person hours / days.
}
\begin{samepage}
\begin{mdframed}[style=MyFrame]
I would estimate that \lang\ has reduced testing effort by up to 80\%. With testing being roughly 20\% of the effort in a typical project, I believe that the tool saves us approximately 15\% of the whole project effort. Based on my involvement in two major projects, this translates to a savings of 3-4 weeks over the past six months.
\end{mdframed}
\end{samepage}

\vspace*{1em}\noindent\textit{\textbf{Q10.} What specific benefits have you observed while using \lang?
}
\begin{samepage}
\begin{mdframed}[style=MyFrame]
Effort savings is obviously one important benefit. I would also say that our stress testing process has become easier to manage because the testing team is not nearly as exposed to network technologies under the hood. Setting up the testing environment, of course, still needs to be done and takes some time, but once we have done that, the rest is just a click.
\end{mdframed}
\end{samepage}

\vspace*{1em}\noindent\textit{\textbf{Q11.} Please rate this statement: \\[.5em]
I found \lang\ to be a useful complement to existing tools for stress testing.\\[.5em]
        \underline{\bfseries$\text{\rlap{$\checkmark$}}\square$Strongly agree},
        $\square$Agree,
        $\square$Neutral,
        $\square$Disagree,
        $\square$Strongly disagree
}

\vspace*{1em}\noindent\textit{\textbf{Q12.} Do you have any additional comments or insights you would like to share regarding \lang? 
}
\begin{samepage}
\begin{mdframed}[style=MyFrame]
The current \lang\ focuses mainly on the transport layer. It would be beneficial to extend its scope to include other layers as well. For example, I think that adding support for RESTful APIs would add value to \lang.
\end{mdframed}
\end{samepage}

\vspace*{-.5em}

\end{appendices}

\end{document}